\documentclass{article}
\usepackage{amsmath,amsthm}
\usepackage{graphicx,epstopdf,epsfig,multirow,epic,bm}
\usepackage{lineno,hyperref}
\usepackage{amsfonts}
\usepackage{subfigure}
\usepackage{amssymb}
\usepackage{algorithm}
\usepackage{algpseudocode}

\normalsize \rm
\begin{document}

\begin{center}
{\Large  \textbf { Randomized Response Mechanisms for Differential Privacy Data Analysis: Bounds and Applications }}\\[12pt]
{\large Fei Ma$^{a,}$\footnote{~The author's E-mail: mafei123987@163.com. },\quad  Ping Wang$^{b,c,d,}$\footnote{~The author's E-mail: pwang@pku.edu.cn.} }\\[6pt]
{\footnotesize $^{a}$ School of Electronics Engineering and Computer Science, Peking University, Beijing 100871, China\\
$^{b}$ National Engineering Research Center for Software Engineering, Peking University, Beijing, China\\
$^{c}$ School of Software and Microelectronics, Peking University, Beijing  102600, China\\
$^{d}$ Key Laboratory of High Confidence Software Technologies (PKU), Ministry of Education, Beijing, China}\\[12pt]
\end{center}

\begin{quote}
\textbf{Abstract:} Randomized response, as a basic building-block for differentially private mechanism, has given rise to great interest and found various potential applications in science communities. In this work, we are concerned with three-elements randomized response (RR$_{3}$) along with relevant applications to the analysis of weighted bipartite graph upon differentially private guarantee. We develop a principled framework for estimating statistics produced by RR$_{3}$-based mechanisms, and then prove the corresponding estimations to be unbiased. At the same time, we study in detail several fundamental and significant members in RR$_{3}$ family, and derive the closed-form solutions to unbiased estimations. Next, we show potential applications of several RR$_{3}$-based mechanisms into the estimation of average degree and average weighted value on weighted bipartite graph when requiring local differential privacy guarantee. In the meantime, we determine the lower bounds for choice of relevant parameters by minimizing variance of statistics in order to design optimal RR$_{3}$-based local differential private mechanisms, with which we optimize previous protocols in the literature and put forward a version that achieves the tight bound. Last but most importantly, we observe that in the analysis of relational data such as weighted bipartite graph, a portion of privacy budget in local differential private mechanism is sometimes ``consumed" by mechanism itself accidentally, resulting to a more stronger privacy guarantee than we would get by simply sequential compositions.
\\

\textbf{Keywords:} Three-elements randomized response, Local differential private mechanism, Weighted bipartite graph, Optimization \\

\end{quote}

\section{Introduction}

Nowadays, with the rapid development of data mining technology, people's daily life has been substantially affected by a great variety of applications upon big data analysis \cite{Salloum-2019,Cuzzocrea-2018}. On the one hand, it provides persons with considerable convenience. For instance, in the context of App developing, service developers can design more friendly Apps for users by collecting and analyzing users' previous data, and further improve the user experience \cite{Apple-2016}-\cite{Ding-2017}. On the other hand, it is well known that the collected data often contain a large amount of privacy information. For example, the health data of users include many highly sensitive information including age, gender, IDs, phone number, address, and disease patterns, and so on. There are no doubt that privacy leakage is inevitable by collecting and analyzing data directly without any security guarantee \cite{Corrigan-Gibbs-2017}. Therefore, it is urgent to put forward available privacy-preserving techniques for the collection and analyses of data. As a promising tool in the realm of data privacy protection, differential privacy has become the de facto standard for private data release due to the fact that it provides provable privacy guarantee in a mathematically rigorous manner \cite{Dwork-2006}-\cite{Duchi-2014}. As such, differential privacy has been widely deployed in various kinds of practical settings including Google \cite{Erlingsson-2014}, Apple \cite{Apple-2016} as well as Microsoft \cite{Ding-2017}, and so forth.

In theory, differential privacy promises privacy protection via injecting noisy into raw data, i.e., perturbing raw data by introducing randomness \cite{Dwork-2006,Dwork-2006-1}. In order to achieve differential privacy, a large number of approaches to generating available randomness have been proposed, such as Laplacian mechanism \cite{Dwork-2006-1}, exponential mechanism \cite{Dwork-2014}, randomized response \cite{Erlingsson-2014}, etc. Among of them, as a simple and well-studied technique, randomized response has attracted increasing attention in the field of differential privacy \cite{Du-2003}-\cite{Gu-2020} in the last several years. As is known to us all, randomized response as a means to estimate bias in a survey is first proposed by Warner in 1965 \cite{Warner-1965}. This technique is initially developed in the social sciences in order to collect statistical quantities about sensitive information, such as embarassing and illegal behavior, that researchers often care about. The original and simplest version of randomized response is for the survey with binary answers. Roundly speaking, the heart is the following: given a piece of sensitive information, denoted by $\xi$, for instance, ``Have you ever cheated in college entrance examination?", study participants are told to report whether or not they have property $\xi$ in the following plausible manner. An individual participant first flips an uneven coin \footnote[1]{Notion ``uneven" indicates that from the statistical point of view, the probability of observing tail (or head), denoted by $p$ (or $q=1-p$), is not $1/2$ when flipping this coin.}, (1) if tail, then he/she sends answer truthfully, (2) else, he/she response opposite answer. Clearly, the motivation for randomized response comes from the plausible deniability of any outcome \cite{Dwork-2014}. As such, participants are willing to engage in survey and truthfully answer what researchers are seeking for. Since Warner proposed randomized response, many researchers have paid more attention to investigation on basic properties of the typical randomized response and its variants. As a result, a great variety of versions regarding to randomized response have been established in the rich literature \cite{Kairouz-2014}-\cite{Nguyen-2016}. Studied models include unrelated question model \cite{Greenberg-1968}, Moor's procedure \cite{Moors-1971}, two-stage model \cite{Mangat-1994}, and so on. These models have been widely adopted into many practical scenarios, for instance, corruption \cite{Gingerich-2010}, sexual behavior \cite{Donovan-2003}, faking on a CV \cite{Chen-2014}.

As we can see above, randomized response is in essence a randomized algorithm. It is natural to ask whether the answers yielded by this technique are true or not. More generally, there are many reasons to suspect whether the outputs of this kind of algorithm are stable or not when supposing that all the results are unbiased. To answer the two concerns above, one needs to appeal to two statistical parameters, expectation and variance, from probability theory and statistic. The former is for the first concern, and the latter is commonly used as a measure for stability of a sequence of random variables. As mentioned previously, the two concerns behind randomized response itself as basic properties have been studied in detail in the literature \cite{Greenberg-1968}-\cite{Chen-2014}. In particular, randomized response, as a basic building-block for differentially private algorithm, has been widely used and, accordingly, the two concerns mentioned above have been analyzed in the specific setting \cite{Wang-2018}-\cite{Gu-2020}. In recently published paper \cite{Holohan-2017}, Holohan \emph{et al} have systematically examined a generalized randomized response in the context of differential privacy. It should be pointed out that Ref.\cite{Holohan-2017} focuses on randomized response with binary answers. As opposed to the previous work, the goal of this work is to study basic properties on randomized response with three answers and then to answer the two concerns above in specific situations. At first sight, this extension is a straightforward consequence. However, as will be shown shortly, it is in fact untrivial mainly due to (1) increasing the number of answers remarkably results in complicated calculations, which makes it more hard to analyze relevant parameters, (2) randomized response of such type is closely connected to the privacy-preserving analysis of relational data such as weighted bipartite graph. In a nutshell, it is of great interest to probe basic properties planted on randomized response with more answers (including three answers).

To summarize, the main contribution of this paper is shown as below. We first establish a principled framework for estimating expectation and variance of statistical quantities output by randomized response with three answers. Then, we focus on several fundamental and significant schemes upon randomized response with three answers, and obtain the closed-form formulas of expectation and variance using the proposed framework. Next, we determine bounds pertaining to relevant parameters in the context of differential privacy, and also show some applications in the analysis of weighted bipartite graph. In the meantime, we provide a guideline to design optimal differentially private algorithm based on randomized response. At last, we list out some interesting and challenging open problems.

\emph{Roadmap}---The rest of this paper is organized as follows. Section II introduces some basic notations used later, such as, local differential privacy, randomized response and bipartite graph. Section III establishes general formulas for estimations and variance of some quantities pertaining to the counting inquiry based on three-elements randomized response \footnote[2]{Throughout this paper, the terms randomized response with three answers and three-elements randomized response are used interchangeably.} by virtue of Maximal Likelihood Estimate. Section IV discusses in detail several classes of fundamental differentially private mechanisms upon three-elements randomized response and shows the corresponding potential applications in the field of analyzing weighted bipartite graph in a privacy-preserving manner. In the meantime, we derive the analytical solutions to bounds for relevant parameters on differentially private mechanisms in question, based on which the protocols achieving tight bound are put forward. Section V briefly reviews the related work. Finally, we close this paper in Section VI.

\section{Notations and Definitions}

In this section, we will introduce some fundamental concepts and notations used later. First of all, let us recall the concrete definition of differential privacy (DP) in the global setting \cite{Dwork-2006}. Specifically, there always exists a trusted party who is assumed to not steal or leak private information of arbitrary data owner in the centralized case. And then, the party serves as a data collector who executes a randomized mechanism on raw data collected from each user and reports the sanitized results for a given query. On the other hand, this assumption is not always reasonable. One of most important reasons for this is that with the rapid development of data mining, data is now deemed as the crucial assets in various kinds of fields. Data owners themselves become more and more concerned with their own data, and worry that they no longer take control of data as long as raw data are sent to that so-called trusted party. As a result, the local version related to differential privacy has been established \cite{Duchi-2013} and also deployed into many practical applications \cite{Wang-2017,Ye-2019,Gu-2020}, \cite{Kairouz-2014}-\cite{Nguyen-2016} where no party but for data owner himself has access to the raw data. More concisely, each data owner perturbs his/her data by using a randomized mechanism which satisfies differential privacy, and then sends the correspondingly sanitized result to the data collector who might be untrusted. The detailed description is shown in Def.1.

\textbf{Definition 1 \cite{Duchi-2013}} A randomized mechanism $\mathcal{M}$ turns out to satisfy \emph{$\epsilon$-local differential privacy}, shorted as $\epsilon$-LDP, if and only if for arbitrary two inputs $D_{i}$ and $D_{j}$ in domain $\mathcal{D}$ and for an arbitrary output $A^{\ast}$ in range $\mathcal{A}$, the following inequality always holds

\begin{equation}\label{eqa:MF-2-1}
\mathrm{Pr}[\mathcal{M}(D_{i})=A^{\ast}]\leq e^{\epsilon}\mathrm{Pr}[\mathcal{M}(D_{j})=A^{\ast}]
\end{equation}
where parameter $\epsilon$ is often called privacy budget that measures the level of privacy-preserving guarantee promised by mechanism $\mathcal{M}$. Roundly speaking, $\epsilon$-LDP mechanism means that after observing the output $A^{\ast}$, an arbitrary adversary including data collector cannot infer whether the input is $D_{i}$ or $D_{j}$ with high confidence controlled by parameter $\epsilon$. A smaller $\epsilon$ implies a stronger privacy-preserving guarantee.

\textbf{Definition 2 \cite{Dwork-2014}} Given a collection of randomized mechanisms $\mathcal{M}_{i}$ ($i\in[1,n]$), each of which satisfies $\epsilon_{i}$-local differential privacy. Then, the sequence of mechanisms $\mathcal{M}_{i}$ together provides $\Sigma_{i=1}^{n}\epsilon_{i}$-local differential privacy. This is the so-called \emph{Sequential Composition} of differential privacy mechanisms.

\emph{Remark 1} The fundamental property defined in Def.2 enables one to design more complicated differential privacy mechanisms using some simple and basic building-blocks. For instance, given a privacy budget $\epsilon$, we can partition it into several smaller portions such that each portion is employed into one randomized mechanism that is collecting useful information from the original data.

\emph{Remark 2} The sequential composition defined in Def.2 always provides an upper for the guarantee of differential privacy. Technically speaking, this demonstration can be interpreted into the coming expression

\begin{equation}\label{eqa:MF-2-2}
\begin{aligned}&\mathfrak{M}=:\left\{\mathcal{M}_{j}^{\star}|\mathcal{M}_{j}^{\star}=:\bigoplus_{j;i=1}^{n}\mathcal{M}_{i},j\in[1,m]\right\}, \\
&\epsilon^{\ddagger}=\sum_{i=1}^{n}\epsilon_{i}=\text{sup}\{\epsilon_{j}^{\star}|j\in[1,m]\}
\end{aligned}
\end{equation}
where $\mathcal{M}_{j}^{\star}$ follows $\epsilon_{j}^{\star}$-local differential privacy, and symbol $\bigoplus_{j;i=1}^{n}$ represents the $j$-th design way between given $n$ local differential privacy mechanisms $\mathcal{M}_{i}$. It is worth noting that, without loss of generality, we here assume that there are $m$ different types of design manners. Based on this, for a given data analysis task and $n$ LDP mechanisms $\mathcal{M}_{i}$, it is of great interest to find the optimal design manners $\mathcal{M}^{\dagger}\in\mathfrak{M}$ achieving the lower bound of privacy budget $\epsilon^{\dagger}=\text{inf}\{\epsilon_{j}^{\star}|j\in[1,m]\}$.

\textbf{Definition 3 \cite{Warner-1965}} For the anticipants in a survey consisting of $m$ potential candidate answers, the so-called \emph{Generalized Randomized Response} (GRR) as a widely-used technique with plausible deniability can be utilized to let a anticipant return a randomized answer to a sensitive question. More precisely, an individual participant sends the genuine answer with probability $p$ or gives an arbitrarily other answer with probability $(1-p)/(m-1)$. For ease of presentation, we call the aforementioned GRR $m$-elements RR. In particular, when $m=2$, we obtain the classic \emph{Randomized Response} (RR) that was firstly proposed by Warner \cite{Warner-1965} in 1965. In this case, we often make use of ``Yes" and ``No" to indicate two distinct kinds of answers.

\emph{Remark 3} The privacy-preserving techniques based on GRR/RR have been employed into some DP/LDP mechanisms including RAPPOR \cite{Erlingsson-2014}, SHist \cite{Bassily-2015}, PrivKVM \cite{Ye-2019}, KVUE \cite{Sun-2019} and PCKV \cite{Gu-2020}. Technically speaking, it is necessary to set probability $p$ to $e^{\epsilon}/(e^{\epsilon}+1)$ for RR to satisfy $\epsilon$-LDP. Similarly, the probability $p$ in GRR should be assumed to be $e^{\epsilon}/(e^{\epsilon}+m-1)$ in order to make GRR to follow $\epsilon$-LDP.

\emph{Remark 4} In general, the percentage of ``Yes" answers (denoted by $f$) directly obtained from RR is bias. Therefore, we need to calibrate the biased result in terms of the following expression
\begin{equation}\label{eqa:MF-2-3}
f'=\frac{f-(1-p)}{p-(1-p)}
\end{equation}
where $f'$ is defined as the corrected proportion that will be reported.

\emph{Remark 5} Mathematically, it is convenient to interpret GRR into a probability matrix $\mathbf{P}_{m\times m}$ each of whose entries $p_{ij}$ represents a probability for inverting answer $i$ into $j$. By definition, matrix $\mathbf{P}_{m\times m}$ is row-stochastic, i.e., $\Sigma_{j=1}^{m}p_{ij}=1$. For case of $m=3$, matrix $\mathbf{P}_{3\times 3}$ is given by

$$\mathbf{P}_{3\times 3}=\left(
               \begin{array}{ccc}
                p_{11} &  p_{12} & p_{13} \\
                   p_{21} &  p_{22} & p_{23} \\
                   p_{31} &  p_{32} & p_{33} \\
               \end{array}
             \right)=(\mathbf{P}^{\top}_{1}, \mathbf{P}^{\top}_{2}, \mathbf{P}^{\top}_{3})$$
in which vector $\mathbf{P}_{i}$ is defined as $\mathbf{P}_{i}=( p_{i1}, p_{i2}, p_{i3})$ and $\top$ indicates transpose. Intuitively, such a representation is closely associated to a directed graph shown in Fig.1. For convenience, this type of interpretation for GRR is always adopted in the rest of this paper.

\begin{figure}
\centering
  \includegraphics[height=6cm]{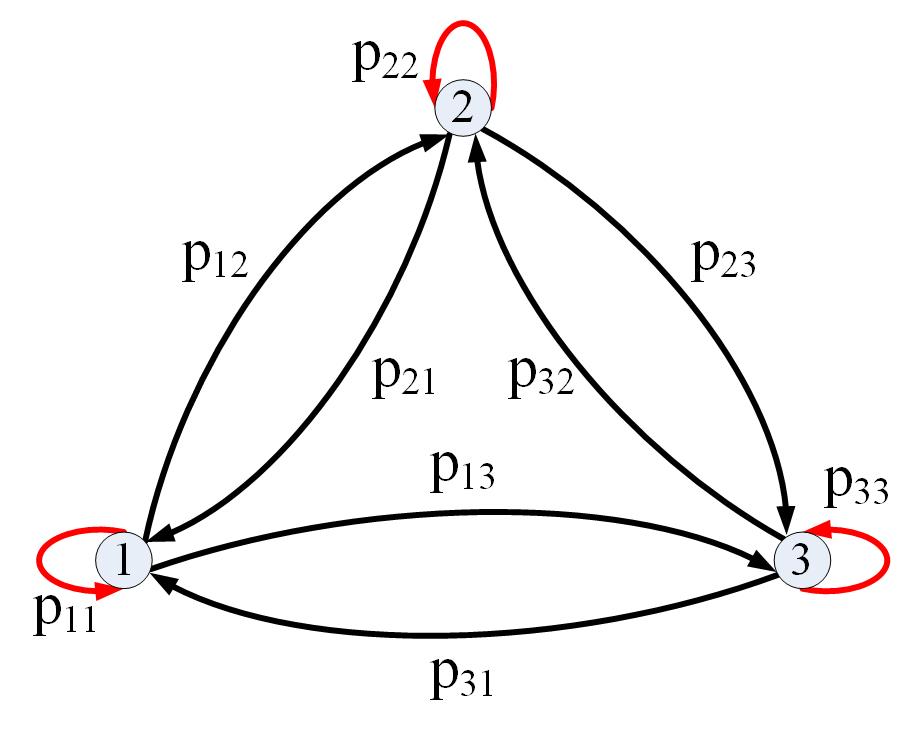}\\
{\small Fig.1. (Color online) The diagram of directed graph corresponding to probability matrix $\mathbf{P}_{3\times 3}$.  }
\end{figure}

\textbf{Definition 4 \cite{Bondy-2008}} It is well known that graph $\mathcal{G}(\mathcal{V},\mathcal{E})$ has proven its power to model relational data in the past years \cite{Barabasi-2016,Newman-2018}. Here, $\mathcal{V}$ and $\mathcal{E}$ represent vertex set and edge set, respectively. For each vertex $v$ in set $\mathcal{V}$, its degree $k_{v}$ is referred to as the number of edges incident with $v$. A graph $\mathcal{G}(\mathcal{V},\mathcal{E})$ is considered bipartite if there is a partition $(\mathcal{V}_{1},\mathcal{V}_{2})$ of vertex set $\mathcal{V}$ so that every edge in $\mathcal{E}$ has one end in $\mathcal{V}_{1}$ and one end in $\mathcal{V}_{2}$. We denote by $\mathcal{K}_{\mathcal{V}_{1},\mathcal{V}_{2}}$ a bipartite graph with partition $(\mathcal{V}_{1},\mathcal{V}_{2})$ for brevity. Let each edge $e$ of $\mathcal{E}$ be associated a real number $w(e)$, called its weight. Then $\mathcal{G}(\mathcal{V},\mathcal{E})$, together with these weights on its edges, is called a weighted graph. In this setting, the weight $w_{v}$ of vertex $v$ in set $\mathcal{V}$ is viewed as the summation of weights on edges incident with $v$. For our purpose, let weight $w(e)$ for each edge all belong to $[-1,1]$.

\emph{Remark 6} In the previous literature of studying LDP, there is a great plenty of work related to bipartite graph $\mathcal{K}_{\mathcal{V}_{1},\mathcal{V}_{2}}$ or its weighted version. For example, one is interested in the estimation about frequency of each item under LDP where all participants (i.e., data owners) are grouped into vertex sub-set $\mathcal{V}_{1}$ and the other vertex sub-set $\mathcal{V}_{2}$ consists of all possible items. In other word, this kind of task aims at estimating degree of each vertex in set $\mathcal{V}_{2}$ in a LDP manner. In general, one cares about weight of each vertex in set $\mathcal{V}_{2}$ when considering estimation about value associated with each item where bipartite graph $\mathcal{K}_{\mathcal{V}_{1},\mathcal{V}_{2}}$ is weighted. Beyond that, the above two cases (i.e., degree and weight estimations) have been considered simultaneously in some published papers \cite{Wei-2020} and thus the relevant schemes satisfying LDP have been built. It should be noted that a slightly different description, namely, Key-Value, is used in \cite{Ye-2019} to depict what we are mentioning.

\emph{Remark 7} A large number of studies associated with graph under LDP have been proceeded over the last decade \cite{Dwork-2014}. Accordingly, two notations, Vertex-LDP and Edge-LDP, have been proposed in order to investigate the underlying structure of graph in a LDP fashion \cite{Raskhodnikova-2016}-\cite{Day-2016}. This is out of the scope of this work. Interested readers refer to \cite{Dwork-2014} for more details.

\textbf{Definition 5} Given a weighted bipartite graph $\mathcal{K}_{\mathcal{V}_{1},\mathcal{V}_{2}}$ mentioned above, we assume that there are $n$ vertices $v^{i}_{1}$ ($i\in\{1,2,\dots,n\}$) in set $\mathcal{V}_{1}$ and $m$ vertices $v^{j}_{2}$ ($j\in\{1,2,\dots,m\}$) in set $\mathcal{V}_{2}$. Without loss of generality, the $i$-vertex $v^{i}_{1}$ is connected to $s_{i}$ vertices $v^{j}_{2}$. To put this another way, there is a collection of edges $\mathcal{E}_{v^{i}_{1}}:=\{e_{v^{i}_{1}v^{j_{l}}_{2}}|l=1,2,\dots,s_{i}\}$ in graph $\mathcal{K}_{\mathcal{V}_{1},\mathcal{V}_{2}}$. Based on this, the degree estimation of vertex $v^{j}_{2}$, denoted by $k_{v^{j}_{2}}$, is thought of as
\begin{equation}\label{eqa:MF-2-4}
k_{v^{j}_{2}}=\left|\left\{v^{i}_{1}|\exists e_{v^{i}_{1}v^{j}_{2}}\in\mathcal{E}_{v^{i}_{1}},\quad i=1,2,\dots, n\right\}\right|
\end{equation}
in which $|\mathcal{X}|$ indicates the cardinality of set $\mathcal{X}$. Analogously, the weight estimation of vertex $v^{j}_{2}$, denoted by $w_{v^{j}_{2}}$, is written as
\begin{equation}\label{eqa:MF-2-5}
w_{v^{j}_{2}}=\sum_{i=1}^{n}\sum_{v^{l}_{2}\in\mathcal{E}_{v^{i}_{1}}}w(e_{v^{i}_{1}v^{l}_{2}})\delta_{l,j}=\sum_{i=1}^{n}w(e_{v^{i}_{1}v^{j}_{2}})
\end{equation}
where symbol $\delta_{l,j}$ is the Kronecker delta function in which $\delta_{l,j}$ is equal to $1$ when $l=j$, and $0$ otherwise. From Eqs.(\ref{eqa:MF-2-4}) and (\ref{eqa:MF-2-5}), the average degree $\langle k\rangle$ and average weighted value $\langle w\rangle$ of weighted bipartite graph $\mathcal{K}_{\mathcal{V}_{1},\mathcal{V}_{2}}$ is respectively expressed as

\begin{equation}\label{eqa:MF-2-5}
\begin{aligned}&\langle k\rangle=\frac{\sum_{v^{j}_{2}\in \mathcal{V}_{2}}k_{v^{j}_{2}}}{(n+m)(n+m-1)/2},\\
 &\langle w\rangle=\frac{\sum_{v^{j}_{2}\in \mathcal{V}_{2}}w_{v^{j}_{2}}}{(n+m)(n+m-1)/2}.
 \end{aligned}
\end{equation}

In general, parameters $n$ and $m$ are assumed to be known in the privacy data analysis. Therefore, the estimations about average degree $\langle k\rangle$ and average weighted value $\langle w\rangle$ on weighted bipartite graph $\mathcal{K}_{\mathcal{V}_{1},\mathcal{V}_{2}}$ can be easily reduced to the estimations of degree $k_{v^{j}_{2}}$ and weight $w_{v^{j}_{2}}$. In another word, we focus mainly on degree $k_{v^{j}_{2}}$ and weight $w_{v^{j}_{2}}$ in the coming discussions.

\emph{Remark 8} In the study of Key-Value in \cite{Ye-2019}, the goal is to measure both frequency of each vertex $v^{j}_{2}$, i.e., $f_{v^{j}_{2}}=k_{v^{j}_{2}}/n$, and mean value related to each vertex $v^{j}_{2}$, namely, $m_{v^{j}_{2}}=w_{v^{j}_{2}}/k_{v^{j}_{2}}$. After some simple algebra, it is clear to see that what we are discussing in weighted bipartite graph $\mathcal{K}_{\mathcal{V}_{1},\mathcal{V}_{2}}$ is similar to what is considered in \cite{Ye-2019}. In a nutshell, degree $k_{v^{j}_{2}}$ and weight $w_{v^{j}_{2}}$ are two basic parameters in weighted bipartite graph, and thus become the focus of this paper.

\emph{Remark 9} In theory, if one can obtain degree estimation relevant to each vertex $v^{i}_{1}$ in set $\mathcal{V}_{1}$ in a LDP way, then the underlying structure of bipartite graph $\mathcal{K}_{\mathcal{V}_{1},\mathcal{V}_{2}}$ is constructed in a LDP manner with respect to Sequential Composition in Def.2. Note that we here make use of results in Def.5. To make further progress, the weighted version can also be created in a similar way if the weighted estimation relevant to each vertex $v^{i}_{1}$ in set $\mathcal{V}_{1}$ is captured using a LDP approach. In a nutshell, this provides a reasonable and effective way to collect relational data which may be modeled using a bipartite graph compared to Vertex-LDP and Edge-LDP \cite{Raskhodnikova-2016}-\cite{Day-2016}. This is left as our next move.

\section{Three-elements Randomized Response}

As mentioned previously, RR as a building-block has been widely used to design a great variety of privacy-preserving schemes satisfying LDP \cite{Duchi-2013}. In particular, the typical RR, i.e., two-elements RR, has been analyzed in detail under LDP situation \cite{Erlingsson-2014,Kairouz-2014,Kairouz-2016,Nguyen-2016,Bassily-2015}. In \cite{Holohan-2017}, Holihan \emph{et al} studied optimal differentially private mechanisms based on both two-elements RR and DP from the theory point of view, and concluded that it is optimal to submit genuine answer with probability $e^{\epsilon}/(e^{\epsilon}+1)$ to data collector. As will be shown shortly, this section aims to study three-elements RR and some relevant properties. At first sight, it seems to be a ``simple" and ``straightforward" extension from two-elements RR. It should be noted that there are indeed some differences between two RRs, which is shown in the following discussions. At the same time, we also list out many applications upon three-elements RR in the next section and, accordingly, demonstrate the importance of mechanisms upon three-elements RR. This further implies that it is not trivial to consider three-elements RR.

\subsection{Estimation}

For convenience and brevity, we make use of symbols $0,1,2$ to represent three distinct elements in three-elements RR. Here, we first assume that there are $N$ participants in total. Then, the number of participants having $0$, $1$ and $2$ is assumed to equal $N_{0}$, $N_{1}$ and $N_{2}$ ($=N-N_{0}-N_{1}$), respectively. Note that we define a vector $\mathbf{N}=(N_{0},N_{1},N_{2})$ for the purpose of computation simplicity as shown shortly. The corresponding proportion is easily calculated to yield $\pi_{0},\pi_{1}$ and $\pi_{2}$. Similarly, they are also collected into a vector $\pi$, namely, $\pi=(\pi_{0},\pi_{1},\pi_{2})$. Now, we are ready to show more details about implementation of three-elements RR.

Specifically, after all the participants honestly send the sanitized answers, which are produced on the basis of probability matrix $\mathbf{P}_{3\times 3}=(p_{ij})$ where $i,j$ is in integer set $\{0,1,2\}$, to a designated data collector, he can derive the proportion $\widetilde{\pi}_{i}$ of element $i$ equal to $0,1$, and $2$ respectively, and obtain a vector $\widetilde{\pi}=(\widetilde{\pi}_{0},\widetilde{\pi}_{1},\widetilde{\pi}_{2})$. It is straightforward to write an expression in vector form as follows

\begin{equation}\label{eqa:MF-3-1-1}
\widetilde{\pi}=\pi\mathbf{P}_{3\times 3}.
\end{equation}

Next, the data collector can without difficulty derive an estimation for the true proportion of each class of participants by performing the following simple calculation

\begin{equation}\label{eqa:MF-3-1-2}
\widehat{\pi}=\widetilde{\pi}\mathbf{P}_{3\times 3}^{-1},
\end{equation}
in which the superscript $-1$ represents the inverse of matrix and vector $\widehat{\pi}$ is of form $(\widehat{\pi}_{0},\widehat{\pi}_{1},\widehat{\pi}_{2})$. Note that we have made use of an assumption that the determinant of matrix $\mathbf{P}_{3\times 3}$, denoted by $\det\mathbf{P}_{3\times 3}$, is not equivalent to zero. Analogously, the following formula is easily obtained

$$\widetilde{\mathbf{N}}=\mathbf{N}\mathbf{P}_{3\times 3},\qquad \widehat{\mathbf{N}}=\widetilde{\mathbf{N}}\mathbf{P}_{3\times 3}^{-1}.$$

\emph{Remark 10} For the special case of each diagonal entry $p_{ii}=1$ in matrix $\mathbf{P}_{3\times 3}$, it is clear to see that all the participants send directly their own genuine answers to data collector in a deterministic manner. That is to say, equalities $\pi=\widetilde{\pi}$ and $\mathbf{N}=\widetilde{\mathbf{N}}$ hold true. More generally, if matrix $\mathbf{P}_{m\times m}$ corresponds to a permutation on $m$ distinct elements, this kind of $m$-elements mechanisms are indeed one of the most classic symmetric cipher models \cite{Rubinstein-Salzedo-2018}. Obviously, these mechanisms considered herein are of deterministic form.

\emph{Remark 11} Due to equality $\pi_{0}+\pi_{1}+\pi_{2}=\widetilde{\pi}_{0}+\widetilde{\pi}_{1}+\widetilde{\pi}_{2}=\widehat{\pi}_{0}+\widehat{\pi}_{1}+\widehat{\pi}_{2}=1$, we need to only determine estimations of parameters $\pi_{0}$ and $\pi_{1}$. Therefore, the following discussions about parameter estimations focus mainly on proportions $\pi_{0}$ and $\pi_{1}$.

\subsection{Expectation and Variance}

By definition, the data collector first collects the sanitized results from all the participants by using three-elements RR. Then, the most fundamental and simplest task is to derive a unbiased estimations for parameters $\pi_{0}$ and $\pi_{1}$. In the language of statistic, an estimation $\widehat{\alpha}$ of parameter $\alpha$ we are discussing in a survey is considered unbiased if the corresponding expectation $E[\widehat{\alpha}]$ converges in probability to parameter $\alpha$ itself. Besides that, one often cares about the level of stability of estimation as well. That is to say, the variance associated with estimation needs to be calculated. In some published works \cite{Yoshikawa-2021,Ogasawara-2015}, one would also like to take advantage of mean square error (MSE) instead of variance to measure the stability of estimation. Unless otherwise noted, the coming relevant demonstrations about stability of estimation are proceeded with respect to variance.

\emph{Theorem 1 Assume that the determinant $\det\mathbf{P}_{3\times3}$ is not equal to zero, the maximum likelihood extimators for $\pi_{0}$ and $\pi_{1}$ are not only a pair of zero roots of the following equation system}

\begin{equation}\label{eqa:MF-3-2-1}
\frac{\partial \ln\mathcal{L}(\pi_{0},\pi_{1})}{\partial\pi_{0}}=0, \quad \frac{\partial \ln\mathcal{L}(\pi_{0},\pi_{1})}{\partial\pi_{1}}=0,
\end{equation}
\emph{in which $\mathcal{L}(\pi_{0},\pi_{1})=P(x_{i}=0)^{N_{0}}P(x_{i}=1)^{N_{1}}[1-P(x_{i}=0)-P(x_{i}=1)]^{N-N_{0}-N_{1}}$, but also unbiased. }

\emph{Proof} Although the coming proof is a straightforward exercise in the standard statistic study, we want to show it for the purpose of making this work more self-contained. The more details are as follows. As mentioned before, there are number $N_{0}$, $N_{1}$ and $N_{2}$ of participants under consideration. From the concrete description of manipulating Maximum Likelihood Extimator (MLE), we can obtain the MLE for a pair of parameters $\pi_{0}$ and $\pi_{1}$,

\begin{equation}\label{eqa:MF-3-2-2}
\begin{aligned}\mathcal{L}(\pi_{0},\pi_{1})&=P(x_{i}=0)^{N_{0}}\times P(x_{i}=1)^{N_{1}}\times [1-P(x_{i}=0)-P(x_{i}=1)]^{N-N_{0}-N_{1}}
\end{aligned}.
\end{equation}

Taking log on both hand sides in Eq.(\ref{eqa:MF-3-2-2}) yields

\begin{equation}\label{eqa:MF-3-2-3}
\begin{aligned}\ln\mathcal{L}&(\pi_{0},\pi_{1})=N_{0}\ln P(x_{i}=0)+N_{1}\ln P(x_{i}=1)+(N-N_{0}-N_{1})\ln[1-P(x_{i}=0)-P(x_{i}=1)]
\end{aligned}.
\end{equation}

Performing derivative on both hand sides in Eq.(\ref{eqa:MF-3-2-3}) with respect to variables $\pi_{0}$ and $\pi_{1}$, respectively, produces

\begin{subequations}
\label{eq:whole}
\begin{eqnarray}
\begin{aligned}\frac{\partial \ln\mathcal{L}}{\partial\pi_{0}}&=\frac{N_{0}+N_{1}-N}{1-P(x_{i}=0)-P(x_{i}=1)}\frac{\partial P(x_{i}=0)}{\partial\pi_{0}}+\frac{N_{0}+N_{1}-N}{1-P(x_{i}=0)-P(x_{i}=1)}\frac{\partial P(x_{i}=1)}{\partial\pi_{0}}\\
&\quad+\frac{N_{0}}{P(x_{i}=0)}\frac{\partial P(x_{i}=0)}{\partial\pi_{0}}+\frac{N_{1}}{P(x_{i}=1)}\frac{\partial P(x_{i}=1)}{\partial\pi_{0}}
\end{aligned},\label{subeq:MF-3-2-4-1}
\end{eqnarray}
\begin{eqnarray}
\begin{aligned}\frac{\partial \ln\mathcal{L}}{\partial\pi_{1}}&=\frac{N_{0}+N_{1}-N}{1-P(x_{i}=0)-P(x_{i}=1)}\frac{\partial P(x_{i}=0)}{\partial\pi_{1}}+\frac{N_{0}+N_{1}-N}{1-P(x_{i}=0)-P(x_{i}=1)}\frac{\partial P(x_{i}=1)}{\partial\pi_{1}}\\
&\qquad+\frac{N_{0}}{P(x_{i}=0)}\frac{\partial P(x_{i}=0)}{\partial\pi_{1}}+\frac{N_{1}}{P(x_{i}=1)}\frac{\partial P(x_{i}=1)}{\partial\pi_{1}}
\end{aligned}.\label{subeq:MF-3-2-4-2}
\end{eqnarray}
\end{subequations}

To make further progress, we can obtain Eqs.(\ref{subeq:MF-3-2-5-1}) and (\ref{subeq:MF-3-2-5-2}). This suggests that the roots corresponding to Eqs.(\ref{subeq:MF-3-2-4-1}) and (\ref{subeq:MF-3-2-4-2}) are what we are trying to seek. In the following, we will prove the MLE for this pair of parameters $\pi_{0}$ and $\pi_{1}$ to be unbiased.

\begin{subequations}
\label{eq:whole}
\begin{eqnarray}
\begin{aligned}\frac{\partial^{2} \ln\mathcal{L}}{\partial\pi^{2}_{0}}&=-\frac{N_{0}}{[P(x_{i}=0)]^{2}}\left(\frac{\partial P(x_{i}=0)}{\partial\pi_{0}}\right)^{2}-\frac{N_{1}}{[P(x_{i}=1)]^{2}}\left(\frac{\partial P(x_{i}=1)}{\partial\pi_{0}}\right)^{2}\\
&\quad-\frac{N-N_{0}-N_{1}}{[1-P(x_{i}=0)-P(x_{i}=1)]^{2}}\left[\frac{\partial P(x_{i}=0)}{\partial\pi_{0}}+\frac{\partial P(x_{i}=1)}{\partial\pi_{0}}\right]^{2}<0
\end{aligned},\label{subeq:MF-3-2-5-1}
\end{eqnarray}
\begin{eqnarray}
\begin{aligned}\frac{\partial^{2} \ln\mathcal{L}}{\partial\pi^{2}_{1}}&=-\frac{N_{0}}{[P(x_{i}=0)]^{2}}\left(\frac{\partial P(x_{i}=0)}{\partial\pi_{1}}\right)^{2}-\frac{N_{1}}{[P(x_{i}=1)]^{2}}\left(\frac{\partial P(x_{i}=1)}{\partial\pi_{1}}\right)^{2}\\
&\quad-\frac{N-N_{0}-N_{1}}{[1-P(x_{i}=0)-P(x_{i}=1)]^{2}}\left[\frac{\partial P(x_{i}=0)}{\partial\pi_{1}}+\frac{\partial P(x_{i}=1)}{\partial\pi_{1}}\right]^{2}<0
\end{aligned}.\label{subeq:MF-3-2-5-2}
\end{eqnarray}
\end{subequations}

As mentioned previously, variable $x_{i}$ is uniformly at random sampled with replacement, and thus is an independent and identically distributed random one. From common sense in probability theory and statistics, we immediately understand $E[N_{i}]=NE[x_{i}]$. To prove MLE for $\pi_{0}$ and $\pi_{1}$ obtained from Eq.(\ref{eqa:MF-3-2-1}) to be unbiased, it suffices to show Eq.(\ref{eqa:MF-3-2-6})
\begin{equation}\label{eqa:MF-3-2-6}
\left\{\begin{aligned}&\begin{aligned}\widehat{\Pi}(\pi_{0},\pi_{1}|\mathbf{P})&=\frac{E[N_{0}](p_{00}-p_{20})}{p_{20}+\pi_{0}(p_{00}-p_{20})+\pi_{1}(p_{10}-p_{20})}-\frac{(p_{02}-p_{22})(N-E[N_{0}]-E[N_{1}])}{p_{22}+\pi_{0}(p_{02}-p_{22})+\pi_{1}(p_{12}-p_{22})}\\
&\quad+\frac{E[N_{1}](p_{01}-p_{21})}{p_{21}+\pi_{0}(p_{01}-p_{21})+\pi_{1}(p_{11}-p_{21})}=0
\end{aligned}\\
&\begin{aligned}\widehat{\Pi}(\pi_{0},\pi_{1}|\mathbf{P})&=\frac{E[N_{1}](p_{10}-p_{20})}{p_{20}+\pi_{0}(p_{00}-p_{20})+\pi_{1}(p_{10}-p_{20})}-\frac{(p_{12}-p_{22})(N-E[N_{0}]-E[N_{1}])}{p_{22}+\pi_{0}(p_{02}-p_{22})+\pi_{1}(p_{12}-p_{22})}\\
&\quad+\frac{E[N_{1}](p_{11}-p_{21})}{p_{21}+\pi_{0}(p_{01}-p_{21})+\pi_{1}(p_{11}-p_{21})}=0
\end{aligned}
\end{aligned}
\right.
\end{equation}
in which

$$E[N_{0}]=NP(x_{i}=0)=N[p_{20}+\pi_{0}(p_{00}-p_{20})+\pi_{1}(p_{10}-p_{20})],$$
and
$$E[N_{1}]=NP(x_{i}=1)=N[p_{21}+\pi_{0}(p_{01}-p_{21})+\pi_{1}(p_{11}-p_{21})].$$
Note that we have used Eq.(\ref{eqa:MF-3-1-1}) in order to derive quantities $P(x_{i}=0)$ and $P(x_{i}=1)$. After some tedious yet simple algebra, Eq.(\ref{eqa:MF-3-2-6}) turns out to be correct. To sum up, this completes the proof of Theorem 1. \qed

In theory, the expression of variance $\text{Var}\left(\widehat{\Pi}(\pi_{0},\pi_{1}|\mathbf{P})\right)$ corresponding to MLE for $\pi_{0}$ and $\pi_{1}$ can be obtained from formulas of MLE for $\pi_{0}$ and $\pi_{1}$. Yet, in view of a fact that the latter is implicitly expressed, it is not easy to derive what we care about. Nonetheless, the thought is straightforward to understand. Taking one monomial $\Gamma N_{i}$ consisting of $N_{i}$ as example, what one needs to do is to replace this term with the corresponding term $\Gamma^{2} \text{Var}(N_{i})$ in form. An ongoing example is shown in Subsection III.C for illustration purpose. In the meantime, the exact solution of $\text{Var}(N_{i})$ is given by

$$\text{Var}(N_{i})=N\text{Var}(x_{j}=i)=N(E[x_{j}^{2}=i]-E[x_{j}=i]^{2})$$
where $E[x_{j}^{2}=i]\equiv E[x_{j}=i]$ for $i=0,1,2$. The precise formula of $E[x_{j}=i]$ is derived from Eq.(\ref{eqa:MF-3-1-1}), such as
$$E[x_{j}=0]=p_{20}+\pi_{0}(p_{00}-p_{20})+\pi_{1}(p_{10}-p_{20})$$

By far, we have discussed three-elements RR in a general setting, and derive some relevant parameters. In what follows, we consider some special cases and use results obtained herein to straightforwardly write formulas for many parameters.

\subsection{Examples}

It is well known that in the typical two-elements RR, namely, Warner's RR, the probability for an arbitrary participant to send genuine answer to data collector is a pre-defined constant. Similarly, a simple version regarding to three-elements RR is obtained by setting all diagonal entries $p_{ii}$ in matrix $\mathbf{P}_{3\times 3}$ in Eq.(\ref{eqa:MF-3-1-1}) to an identical value. Based on this, we can further build up the simplest version of three-elements RR, which corresponds to the following matrix

$$\mathbf{P}_{3\times 3}=\left(
               \begin{array}{ccc}
                p &  (1-p)/2 & (1-p)/2  \\
                   (1-p)/2  &  p & (1-p)/2  \\
                   (1-p)/2  &  (1-p)/2  & p \\
               \end{array}
             \right).$$
For convenience, this version is called Extended Warner's RR (EWRR$_{3}$).

\emph{Corollary 1 Suppose that $p\neq1/3$, then maximum likelihood extimators MLE for parameters $\pi_{0}$ and $\pi_{1}$ of EWRR$_{3}$ are given by}

\begin{equation}\label{eqa:MF-3-3-1}
\left\{\begin{aligned}&\widehat{\Pi}(\pi_{0}|\mathbf{P})=\frac{(1-p)(N-2N_{0})-2pN_{0}}{(1-3p)N}\\
&\widehat{\Pi}(\pi_{1}|\mathbf{P})=\frac{(1-p)(N-2N_{1})-2pN_{1}}{(1-3p)N}
\end{aligned}
\right..
\end{equation}
\emph{Simultaneously, the MLE derived is unbiased and the corresponding variance is given by}

\begin{equation}\label{eqa:MF-3-3-2}
\left\{\begin{aligned}&\text{Var}\left(\widehat{\Pi}(\pi_{0}|\mathbf{P})\right)=\frac{1-p^{2}+2p(3p-1)\pi_{0}-(3p-1)^{2}\pi^{2}_{0}}{(3p-1)^{2}N}\\
&\text{Var}\left(\widehat{\Pi}(\pi_{1}|\mathbf{P})\right)=\frac{1-p^{2}+2p(3p-1)\pi_{1}-(3p-1)^{2}\pi^{2}_{1}}{(3p-1)^{2}N}
\end{aligned}
\right..
\end{equation}

\emph{Proof 1} First of all, from Eqs.(\ref{eqa:MF-3-3-1}) and (\ref{eqa:MF-3-3-2}), we see that it suffices to study quantities $\widehat{\Pi}(\pi_{0}|\mathbf{P})$ and $\text{Var}\left(\widehat{\Pi}(\pi_{0}|\mathbf{P})\right)$ by symmetry. In principle, it is straightforwardly accomplished in a similar manner adopted in proof of Theorem 1. On the other hand, it is clear to the eye that more detailed calculations are not contained in proof of Theorem 1. Taking into account demonstrations above, we attempt to provide a brief proof for the purpose of making all the calculations more concrete and concise. Note that the key thought is based on proof of Theorem 1.

(1) Establishing MLE for parameters $\pi_{0}$ and $\pi_{1}$, denoted by $\widehat{\Pi}(\pi_{0},\pi_{1}|\mathbf{P})$,

\begin{equation}\label{eqa:MF-3-3-3}
\begin{aligned}\widehat{\Pi}(\pi_{0},\pi_{1}|\mathbf{P})&=P(x_{i}=0)^{N_{0}}\times P(x_{i}=1)^{N_{1}}\\
&\quad\times[1-P(x_{i}=0)-P(x_{i}=1)]^{N-N_{0}-N_{1}}
\end{aligned}\end{equation}

(2) Taking log of MLE $\widehat{\Pi}(\pi_{0},\pi_{1}|\mathbf{P})$,

\begin{equation}\label{eqa:MF-3-3-4}
\begin{aligned}\ln&\widehat{\Pi}(\pi_{0},\pi_{1}|\mathbf{P})=N_{0}\ln P(x_{i}=0)+N_{1}\ln P(x_{i}=1)\\
&\quad+(N-N_{0}-N_{1})\ln[1-P(x_{i}=0)-P(x_{i}=1)]
\end{aligned}
\end{equation}

(3) Performing derivative on both hand sides of Eq.(\ref{eqa:MF-3-3-4}) with respect to $\pi_{0}$ and $\pi_{1}$, respectively,

\begin{subequations}
\label{eq:whole}
\begin{eqnarray}
\begin{aligned}\frac{\partial\ln\widehat{\Pi}(\pi_{0},\pi_{1}|\mathbf{P})}{\partial\pi_{0}}&=\frac{(3p-1)N_{0}}{(1-p)+(3p-1)\pi_{0}}+\frac{(1-3p)(N-N_{0}-N_{1})}{2p+(1-3p)\pi_{0}+(1-3p)\pi_{1}}
\end{aligned}
,\label{subeq:MF-3-3-5-1}
\end{eqnarray}
\begin{eqnarray}
\begin{aligned}\frac{\partial\ln\widehat{\Pi}(\pi_{0},\pi_{1}|\mathbf{P})}{\partial\pi_{1}}&=\frac{(3p-1)N_{1}}{(1-p)+(3p-1)\pi_{1}}+\frac{(1-3p)(N-N_{0}-N_{1})}{2p+(1-3p)\pi_{0}+(1-3p)\pi_{1}}
\end{aligned}.\label{subeq:MF-3-3-5-2}
\end{eqnarray}
\end{subequations}

(4) Setting $\frac{\partial\ln\widehat{\Pi}(\pi_{0},\pi_{1}|\mathbf{P})}{\partial\pi_{0}}=0$ and $\frac{\partial\ln\widehat{\Pi}(\pi_{0},\pi_{1}|\mathbf{P})}{\partial\pi_{1}}=0$, then solving for parameters $\pi_{0}$ and $\pi_{1}$ yields results shown in Eq.(\ref{eqa:MF-3-3-1}).

Next, let us pay attention to consolidation of unbiasedness of MLE $\widehat{\Pi}(\pi_{0},\pi_{1}|\mathbf{P})$. As mentioned above, we just consider quantity $\widehat{\Pi}(\pi_{0}|\mathbf{P})$. From the corresponding expression in Eq.(\ref{eqa:MF-3-3-1}), we derive

\begin{equation}\label{eqa:MF-3-3-6}
\begin{aligned}E\left[\widehat{\Pi}(\pi_{0}|\mathbf{P})\right]&=\frac{(1-p)(N-2E[N_{0}])-2pE[N_{0}]}{(1-3p)N}\\
&=\frac{(1-p)N-2NE[x_{i}=0]}{(1-3p)N}\\
&=\frac{(1-p)-2\left[\frac{1-p}{2}+\frac{3p-1}{2}\pi_{0}\right]}{1-3p}\\
&=\pi_{0}
\end{aligned}.
\end{equation}
We omit verification of $\widehat{\Pi}(\pi_{1}|\mathbf{P})$ due to a similar analysis as above.

Lastly, we discuss about Variance of MLE $\widehat{\Pi}(\pi_{0},\pi_{1}|\mathbf{P})$. Analogously, only $\text{Var}\left(\widehat{\Pi}(\pi_{0}|\mathbf{P})\right)$ is studied in detail, as follows

\begin{equation}\label{eqa:MF-3-3-7}
\begin{aligned}\text{Var}\left[\widehat{\Pi}(\pi_{0}|\mathbf{P})\right]&=\text{Var}\left[\frac{(1-p)(N-2N_{0})-2pN_{0}}{(1-3p)N}\right]\\
&=\frac{4\text{Var}[N_{0}]}{(1-3p)^{2}N^{2}}\\
&=\frac{4\left[\frac{1-p}{2}+\frac{3p-1}{2}\pi_{0}\right]\left(1-\left[\frac{1-p}{2}+\frac{3p-1}{2}\pi_{0}\right]\right)}{(1-3p)^{2}N}\\
&=\frac{1-p^{2}+2p(3p-1)\pi_{0}-(3p-1)^{2}\pi^{2}_{0}}{(3p-1)^{2}N}
\end{aligned},
\end{equation}
in which we have used

$$\begin{aligned}&\text{Var}[N_{0}]=N\text{Var}[x_{i}=0],\\
&\text{Var}[x_{i}=0]=E[x^{2}_{i}=0]-E^{2}[x_{i}=0].
\end{aligned}$$

To sum up, this completes the proof of Corollary 1. \qed

As stated above, it is necessary for us to consider a system of linear equations composed of two equations in order to obtain MLE $\widehat{\Pi}(\pi_{i}|\mathbf{P})$. On the other hand, when the number of equations under consideration is increasing, this method become tedious. In this case considered herein, however, there is a more simple method for solving parameters we are interested in. In what follows, let us sketch out another simple proof.

\emph{Proof 2} According to an assumption that each elements $x_{i}$ is an independent and identically distributed random variable, the probability matrix $\mathbf{P}_{3\times 3}$ is converted into the following form

$$\mathbf{P}_{2\times 2}=\left(
               \begin{array}{cc}
                p &  1-p \\
                   (1-p)/2  & (1+p)/2
               \end{array}
             \right).$$
More specifically, all the participants are now only grouped into two classes: (1) Class $A$ contains participants whose elements are $0$, and other participants are collected into Class $B$. Without loss of generality, we assume that each participant in Class $B$ has element $1$. Then, each participant in Class $A$ reports answer $0$ with probability $p$ and answer $1$ with probability $(1-p)$. A similar explanation is suitable for each participant in Class $B$ where the corresponding probability correspond to each entry of the second row in matrix $\mathbf{P}_{2\times 2}$, respectively. Next, by using a similar calculation as shown in proof 1, we need to perform:

(1) Establishing MLE $\widehat{\Pi}(\pi_{0}|\mathbf{P})$,

\begin{equation}\label{eqa:MF-3-3-3-1}
\widehat{\Pi}(\pi_{0}|\mathbf{P})=P(x_{i}=0)^{N_{0}}P(x_{i}=1)^{N-N_{0}}
\end{equation}

(2) Taking log of MLE $\widehat{\Pi}(\pi_{0}|\mathbf{P})$,

\begin{equation}\label{eqa:MF-3-3-4-1}
\ln\widehat{\Pi}(\pi_{0}|\mathbf{P})=N_{0}\ln P(x_{i}=0)+(N-N_{0})\ln P(x_{i}=1)
\end{equation}

(3) Performing derivative on both hand sides of Eq.(\ref{eqa:MF-3-3-4-1}) with respect to $\pi_{0}$ and $\pi_{1}$, respectively,

\begin{equation}\label{eqa:MF-3-3-5-1}
\frac{\partial\ln\widehat{\Pi}(\pi_{0}|\mathbf{P})}{\partial\pi_{0}}=\frac{(3p-1)N_{0}}{(1-p)+(3p-1)\pi_{0}}+\frac{(1-3p)(N-N_{0})}{1+p+(1-3p)\pi_{0}}
\end{equation}

(4) Setting $\frac{\partial\ln\widehat{\Pi}(\pi_{0}|\mathbf{P})}{\partial\pi_{0}}=0$, then solving for parameter $\pi_{0}$  yields

\begin{equation}\label{eqa:MF-3-3-6-1}
\widehat{\Pi}(\pi_{0}|\mathbf{P})=\frac{1-p}{1-3p}-\frac{2N_{0}}{(1-3p)N}
\end{equation}
as desired. At last, it is easy to obtain MLE $\widehat{\Pi}(\pi_{1}|\mathbf{P})$ by symmetry. \qed

Clearly, we obtain what we care about only by solving an equation. This suggests that the method used in proof 2 is more manageable compared to that used in proof 1. To make further progress, we derive a more general formula using method in proof 2 as will be shown shortly.

In general, we define the EWRR$_{n}$ with $n$-elements as follows. In the corresponding probability matrix $\mathbf{P}_{n\times n}$, each diagonal entry $p_{ii}$ is equal to $p$ and all non-diagonal entry $p_{ij}$ is equivalent to $(1-p)/(n-1)$. Using a similar calculation way in proof 2, we come to the following corollary.

\emph{Corollary 2 Suppose that $p\neq1/n$, then maximum likelihood extimators MLE for parameters $\pi_{0},\pi_{1},\cdots,\pi_{n-2}$ of EWRR$_{n}$ are given by}

\begin{equation}\label{eqa:MF-3-3-8}
\widehat{\Pi}(\pi_{i}|\mathbf{P})=\frac{(1-p)[N-(n-1)N_{i}]-(n-1)pN_{i}}{(1-np)N}
\end{equation}
\emph{Simultaneously, the MLE derived is unbiased and the corresponding variance is given by}

\begin{equation}\label{eqa:MF-3-3-9}
\text{Var}\left(\widehat{\Pi}(\pi_{0}|\mathbf{P})\right)=\frac{[1-p+(np-1)\pi_{0}][p-(np-1)\pi_{0}]}{(np-1)^{2}N}.
\end{equation}

In fact, the versions regarding to three-elements RRs are quite rich compared to that composed of two-elements RR. By far, we only study the simplest version, i.e., EWRR$_{3}$. As an extension of EWRR$_{3}$, we are going to analyze another three simple yet useful three-elements RRs. For convenience, we call them RR$_{3}^{\dag}$, RR$_{3}^{\ddag}$ and RR$_{3}^{\clubsuit}$. Note that one can be able to use a similar technique used in previous subsections to discuss these three-elements RRs, and we thus omit the concrete proofs for the sake of simplicity. It should also be mentioned that the great usefulness of these three-elements RRs, that is to say, the deployment in privacy preserving data analysis, is deferred to show in next section.

From now on, let us first study RR$_{3}^{\dag}$. As before, the first step is to describe the associated probability matrix $\mathbf{P}^{\dag}_{3\times 3}$, which is defined as below

$$\mathbf{P}^{\dag}_{3\times 3}=\left(
               \begin{array}{ccc}
                p_{1} &  (1-p_{1})/2 & (1-p_{1})/2  \\
                   (1-p_{2})/2  &  p_{2} & (1-p_{2})/2  \\
                   (1-p_{2})/2 &  (1-p_{2})/2 & p_{2} \\
               \end{array}
             \right).$$
Compared with the above-mentioned matrix $\mathbf{P}_{3\times 3}$, the only difference is that there are two parameters $p_{1}$ and $p_{2}$ contained in matrix $\mathbf{P}^{\dag}_{3\times 3}$. Then, we take advantage of a similar analysis as in Corollary 2 to obtain what we want, which is shown in Corollary 3.

\emph{Corollary 3 Suppose that $p_{2}+2p_{1}-1\neq0$ and $p_{2}\neq1/3$, then maximum likelihood extimators MLE for parameters $\pi_{0},\pi_{1}$ of RR$_{3}^{\dag}$ are written in the following form}

\begin{equation}\label{eqa:MF-3-3-110}
\left\{\begin{aligned}&\widehat{\Pi}(\pi_{0}|\mathbf{P}^{\dag})=\frac{p_{2}-1}{p_{2}+2p_{1}-1}+\frac{2N_{0}}{(p_{2}+2p_{1}-1)N}\\
&\begin{aligned}\widehat{\Pi}(\pi_{1}|\mathbf{P}^{\dag})&=\frac{p_{1}-1}{p_{2}+2p_{1}-1}-\frac{1}{3p_{2}-1}\\
&\quad-\frac{N_{0}}{(p_{2}+2p_{1}-1)N}+\frac{2N_{1}+N_{0}}{(3p_{2}-1)N}
\end{aligned}
\end{aligned}
\right..
\end{equation}
\emph{Simultaneously, the MLEs derived are unbiased and the corresponding variance is given by}

\begin{equation}\label{eqa:MF-3-3-111}
\text{Var}\left(\widehat{\Pi}(\pi_{0}|\mathbf{P}^{\dag})\right)=\frac{(1-p^{2}_{2})+2p_{2}(2p_{1}+p_{2}-1)\pi_{0}-(2p_{1}+p_{2}-1)^{2}\pi^{2}_{0}}{(2p_{1}+p_{2}-1)^{2}N}.
\end{equation}

Furthermore, following the similar research line as above, we can continue to modify matrix $\mathbf{P}_{3\times 3}$ to create other versions using many other tricks of interest, such as, tuning parameter. As a case study, we put forward RR$_{3}^{\ddag}$. First of all, we define the associated probability matrix $\mathbf{P}^{\ddag}_{3\times 3}$

$$\mathbf{P}^{\ddag}_{3\times 3}=\left(
               \begin{array}{ccc}
                p_{1} &  (1-p_{1})/2 & (1-p_{1})/2  \\
                   1-p_{1}  &  p_{1}p_{2} & p_{1}(1-p_{2})  \\
                   1-p_{1} &  p_{1}(1-p_{2})  & p_{1}p_{2} \\
               \end{array}
             \right).$$
It is easy to see that two remarkable difference between matrices $\mathbf{P}^{\ddag}_{3\times 3}$ and $\mathbf{P}^{\dag}_{3\times 3}$ are that (1) the probability for anticipant with answer $1$ or $2$ to report genuine answer is subject to two different parameters, namely, $p_{1}$ and $p_{2}$, and (2) the probability for anticipant with answer $1$ or $2$ to report either other answer is no longer an identical value when $p_{1}(2-p_{2})\neq1$. As will see in Section IV, according to these remarkable differences, RR$_{3}^{\ddag}$ has been used as a basic ingredient of a scheme for studing weighted bipartite graph in a privacy preserving manner \cite{Dwork-2014}. Now, we demonstrate some fundamental properties of RR$_{3}^{\ddag}$.

\emph{Corollary 4 Suppose that $p_{1}\neq1/2$ and $p_{1}p_{2}\neq1/2$, then maximum likelihood extimators MLE for parameters $\pi_{0},\pi_{1}$ of RR$_{3}^{\ddag}$ are written in the following form}

\begin{equation}\label{eqa:MF-3-3-120}
\left\{\begin{aligned}&\widehat{\Pi}(\pi_{0}|\mathbf{P}^{\ddag})=\frac{p_{1}-1}{2p_{1}-1}+\frac{N_{0}}{(2p_{1}-1)N}\\
&\begin{aligned}\widehat{\Pi}(\pi_{1}|\mathbf{P}^{\ddag})&=\frac{p_{1}}{2p_{1}-1}-\frac{1}{2p_{1}p_{2}-1}\\
&\quad-\frac{N_{0}}{(2p_{1}-1)N}+\frac{2N_{1}+N_{0}}{(2p_{1}p_{2}-1)N}
\end{aligned}
\end{aligned}
\right..
\end{equation}
\emph{Simultaneously, the MLE derived is unbiased and the corresponding variance is given by}

\begin{equation}\label{eqa:MF-3-3-121}
\text{Var}\left(\widehat{\Pi}(\pi_{0}|\mathbf{P}^{\ddag})\right)=\frac{p_{1}(1-p_{1})+(2p_{1}-1)^{2}\pi_{0}-(2p_{1}-1)^{2}\pi^{2}_{0}}{(2p_{1}-1)^{2}N}.
\end{equation}

Subsequently, if we introduce more parameters to adjust matrix $\mathbf{P}_{3\times 3}$, then more complicated versions are developed. For instance, we take three parameters $p_{1}$, $p_{2}$ and $q$ to come up with the probability matrix $\mathbf{P}^{\clubsuit}_{3\times 3}$

$$\mathbf{P}^{\clubsuit}_{3\times 3}=\left(
               \begin{array}{ccc}
                p_{1} &  (1-p_{1})/2 & (1-p_{1})/2  \\
                   1-p_{2}  &  p_{2}q & p_{2}(1-q)  \\
                   1-p_{2} &  p_{2}(1-q)  & p_{2}q \\
               \end{array}
             \right).$$
As a result, the three-elements RR based on matrix $\mathbf{P}^{\clubsuit}_{3\times 3}$ is called RR$_{3}^{\clubsuit}$. More detailed applications of RR$_{3}^{\clubsuit}$ are discussed in next section. Before then, with an in sprit similar calculation as above, we come to Corollary 5.

\emph{Corollary 5 Suppose that $p_{2}+p_{1}-1\neq0$ and $q\neq1/2$, then maximum likelihood extimators MLE for parameters $\pi_{0},\pi_{1}$ of RR$_{3}^{\clubsuit}$ are written in the following form}

\begin{equation}\label{eqa:MF-3-3-130}
\left\{\begin{aligned}&\widehat{\Pi}(\pi_{0}|\mathbf{P}^{\clubsuit})=\frac{p_{2}-1}{p_{2}+p_{1}-1}+\frac{N_{0}}{(p_{2}+p_{1}-1)N}\\
&\begin{aligned}\widehat{\Pi}(\pi_{1}|\mathbf{P}^{\clubsuit})&=\frac{p_{1}}{2(p_{2}+p_{1}-1)}-\frac{1}{2p_{2}(2q-1)}\\
&\quad-\frac{N_{0}}{2(p_{2}+p_{1}-1)N}+\frac{2N_{1}+N_{0}}{2p_{2}(2q-1)N}
\end{aligned}
\end{aligned}
\right..
\end{equation}
\emph{Simultaneously, the MLE derived is unbiased and the corresponding variance is given by}

\begin{equation}\label{eqa:MF-3-3-131}
\text{Var}\left(\widehat{\Pi}(\pi_{0}|\mathbf{P}^{\clubsuit})\right)=\frac{p_{2}(1-p_{2})+(2p_{2}-1)(p_{1}+p_{2}-1)\pi_{0}-(p_{1}+p_{2}-1)^{2}\pi^{2}_{0}}{(p_{1}+p_{2}-1)^{2}N}.
\end{equation}

By far, we obtain general formulas of parameter estimation in three-elements RRs using maximum likelihood extimator. In particular, several simple and fundamental members in three-elements RRs are selected to serve as examples to show concrete computational manipulations. As known, RR is one of best-used building-blocks for designing DP mechanisms, and has given rise to a high degree of academic attention \cite{Kairouz-2014}-\cite{Nguyen-2016}. In the following section, we will elaborate on applications of RRs mentioned above to mechanisms guaranteeing privacy preserving data analysis, especially, in the setting of weighted bipartite graph.

\section{Bounds and applications}

The goal of this section is to use the aforementioned three-elements RRs to establish privacy preserving mechanisms complying with LDP, and to find as optimal as possible mechanism by minimizing the corresponding variance. In addition, we also analyze two published LDP protocols that are suitable for studying weighted bipartite graph. In a word, this section is divided into two parts. The first part aims to optimize several three-elements RRs under LDP situation, and three practical applications are discussed in depth in the second part.

\subsection{Bounds}

For an optimization problem, it is key to first determine the feasibility region. As mentioned previously, we aim to optimize some three-elements RRs subject to LDP. That is to say, the feasibility region associated with the following optimization problems is based on requirements from LDP itself. As the first example, let us convert insight into discussion about the simplest three-elements RR, i.e., EWRR$_{3}$.

\subsubsection{EWRR$_{3}$}

By definition, EWRR$_{3}$ is built upon matrix $\mathbf{P}_{3\times 3}$ where each diagonal entry is assumed to be the same value $p$, and all no-diagonal entries are also assumed to equal an identical value $(1-p)/2$. As opposed to the typical Warner's RR, we need to consider two different cases according to probability $p$ as below.

\textbf{Case 1.} In this case, we require that probability $p$ be no less than $1/3$. At the same time, in order to make EWRR$_{3}$ satisfy LDP, it is clear to see from Def.1 that probability $p$ meets an inequality $2p/(1-p)\leq e^{\epsilon}$. To sum up, the feasibility region corresponding to EWRR$_{3}$, denoted by $\Omega_{3}$, is as follows

$$\Omega_{3}:=\left\{p| \frac{1}{3}<p<1,\frac{2p}{1-p}\leq e^{\epsilon}\right\}=\left\{p| \frac{1}{3}<p\leq \frac{e^{\epsilon}}{e^{\epsilon}+2}\right\}.$$
Then, the central question is to find

$$\arg\min_{p\in\Omega_{3}}\text{Var}\left(\widehat{\Pi}(\pi_{0}|\mathbf{P})\right).$$

\emph{Theorem 2 For EWRR$_{3}$ where $p>1/3$, we have }

\begin{equation}\label{eqa:MF-4-1-1}
\arg\min_{p\in\Omega_{3}}\text{Var}\left(\widehat{\Pi}(\pi_{0}|\mathbf{P})\right)=\left\{\frac{e^{\epsilon}}{e^{\epsilon}+2}\right\}.
\end{equation}

\emph{Proof} While the following proof is straightforwardly proceeded using method in any standard mathematics book-text, we include it in the interest of completeness.

Performing derivative of quantity $\text{Var}\left(\widehat{\Pi}(\pi_{0}|\mathbf{P})\right)$ with respect to $p$ yields
\begin{equation}\label{eqa:MF-4-1-2}
\frac{\partial\text{Var}\left(\widehat{\Pi}(\pi_{0}|\mathbf{P})\right)}{\partial p}=\frac{2p-6-2(3p-1)\pi_{0}}{(3p-1)^{3}N}<0.
\end{equation}
This implies that $\text{Var}\left(\widehat{\Pi}(\pi_{0}|\mathbf{P})\right)$ is a decreasing function over variable $p$. So, the minimum is attained at point $p=\frac{e^{\epsilon}}{e^{\epsilon}+2}$. Therefore, we complete the proof of Theorem 2. \qed

\textbf{Case 2.} Here, we require that probability $p$ be less than $1/3$. Now, in order to make EWRR$_{3}$ satisfy LDP, it is easy to see from Def.1 that probability $p$ obeys an inequality $(1-p)/2p\leq e^{\epsilon}$. Taken together, the feasibility region corresponding to EWRR$_{3}$, denoted by $\Omega'_{3}$, is as follows

$$\Omega'_{3}:=\left\{p|0<p<\frac{1}{3},\frac{1-p}{2p}\leq e^{\epsilon}\right\}=\left\{p| \frac{1}{2e^{\epsilon}+1}\leq p<\frac{1}{3} \right\}.$$
Next, the key question is to seek

$$\arg\min_{p\in\Omega'_{3}}\text{Var}\left(\widehat{\Pi}(\pi_{0}|\mathbf{P})\right).$$

\emph{Theorem 3 For EWRR$_{3}$ where $p<1/3$, we have }

\begin{equation}\label{eqa:MF-4-1-3}
\arg\min_{p\in\Omega'_{3}}\text{Var}\left(\widehat{\Pi}(\pi_{0}|\mathbf{P})\right)=\left\{\frac{1}{2e^{\epsilon}+1}\right\}.
\end{equation}

\emph{Proof} From Eq.(\ref{eqa:MF-4-1-2}), we see

\begin{equation}\label{eqa:MF-4-1-4}
\frac{\partial\text{Var}\left(\widehat{\Pi}(\pi_{0}|\mathbf{P})\right)}{\partial p}=\frac{2p-6-2(3p-1)\pi_{0}}{(3p-1)^{3}N}>0.
\end{equation}
This implies that in feasibility region $\Omega'_{3}$, $\text{Var}\left(\widehat{\Pi}(\pi_{0}|\mathbf{P})\right)$ is now an increasing function over variable $p$. Therefore, the minimum is attained at point $p=\frac{1}{2e^{\epsilon}+1}$. This completes the proof of Theorem 3. \qed

In general, we consider EWRR$_{n}$, and, after using a similar analysis as above, conclude

\begin{subequations}
\label{eq:whole}
\begin{eqnarray}
\arg\min_{p\in\Omega_{n}}\text{Var}\left(\widehat{\Pi}(\pi_{0}|\mathbf{P})\right)=\left\{\frac{e^{\epsilon}}{e^{\epsilon}+n-1}\right\},\label{subeq:MF-4-1-5-1}
\end{eqnarray}
\begin{eqnarray}
\arg\min_{p\in\Omega'_{n}}\text{Var}\left(\widehat{\Pi}(\pi_{0}|\mathbf{P})\right)=\left\{\frac{1}{(n-1)e^{\epsilon}+1}\right\}.\label{subeq:MF-4-1-5-2}
\end{eqnarray}
\end{subequations}
in which symbols $\Omega_{n}$ and $\Omega'_{n}$ indicate the feasibility regions satisfied by EWRR$_{n}$ in two different settings, respectively, and are defined as

$$\begin{aligned}&\Omega_{n}:=\left\{p| \frac{1}{n}<p\leq \frac{e^{\epsilon}}{e^{\epsilon}+n-1}\right\},\\
&\Omega'_{n}:=\left\{p| \frac{1}{(n-1)e^{\epsilon}+1}\leq p<\frac{1}{n}\right\}
\end{aligned}$$

It should be pointed out that parameter $\epsilon$ is often assumed to be more than $0$. Clearly, if parameter $\epsilon$ is equal to $0$, then the feasibility region reduces into a point $1/n$. In this sense, each participant sends uniformly at random an arbitrary answer in anticipated elements space to the data collector. To put this another way, random variable $x_{i}$ follows uniform distribution, which leads the data collector to obtain nothing about original data what he is interested in. This further suggests that such a survey is meaningless while the strongest privacy-preserving guarantee is achieved. Thereby, without loss of generality, we assume that in the following discussions, the probability for participant to send genuine answer is no less than that of sending either other anticipated answer. Note that other cases can be analyzed in an in spirit similar manner, thus we omit them for the sake of simplicity.

\subsubsection{RR$^{\dag}_{3}$}

In this subsection, we aim at considering RR$^{\dag}_{3}$ under the constraint of LDP. In line with the analysis above, the first step is to ascertain the feasibility region $\Omega^{\dag}_{3}$. First of all, let $p_{1}$ not be equal to $p_{2}$ because the case of $p_{1}=p_{2}$ has been probed in subsection IV.A.1. Then, in view of demonstrations above, $\Omega^{\dag}_{3}$ is evaluated to produce

$$\Omega^{\dag}_{3}:=\left\{(p_{1},p_{2})|\quad\begin{aligned} &\frac{1}{3}<p_{1}<1,\; \frac{1}{3}<p_{2}<1\\
&2p_{1}\leq e^{\epsilon}(1-p_{2})\\
&2p_{2}\leq e^{\epsilon}(1-p_{1})\\
&2p_{2}\leq e^{\epsilon}(1-p_{2})
\end{aligned}\right\}.$$
Due to the last two inequalities, we should consider two cases as follows

$$\begin{aligned}&\Omega^{\dag}_{3}(1):=\left\{(p_{1},p_{2})|\quad\begin{aligned} &\frac{1}{3}<p_{1}<1\\
&p_{1}<p_{2}\\
&2p_{1}\leq e^{\epsilon}(1-p_{2})\\
&\frac{1}{3}<p_{2}\leq \frac{e^{\epsilon}}{e^{\epsilon}+2}
\end{aligned}\right\}, \quad \text{and}\\
&\Omega^{\dag}_{3}(2):=\left\{(p_{1},p_{2})|\quad\begin{aligned} &\frac{1}{3}<p_{1}<1,\; \frac{1}{3}<p_{2}<1\\
&p_{2}<p_{1}\\
&2p_{1}\leq e^{\epsilon}(1-p_{2})\\
&2p_{2}\leq e^{\epsilon}(1-p_{1})
\end{aligned}\right\}
\end{aligned}$$
To make our demonstration more concrete, the graphic outlines of the feasibility regions $\Omega^{\dag}_{3}(1)$ and $\Omega^{\dag}_{3}(2)$ are plotted in Fig.2.

\begin{figure*}
\centering
  \centering
\subfigure[$\Omega^{\dag}_{3}(1)$]{
\begin{minipage}[t]{0.4\linewidth}
\centering
\includegraphics[width=6cm]{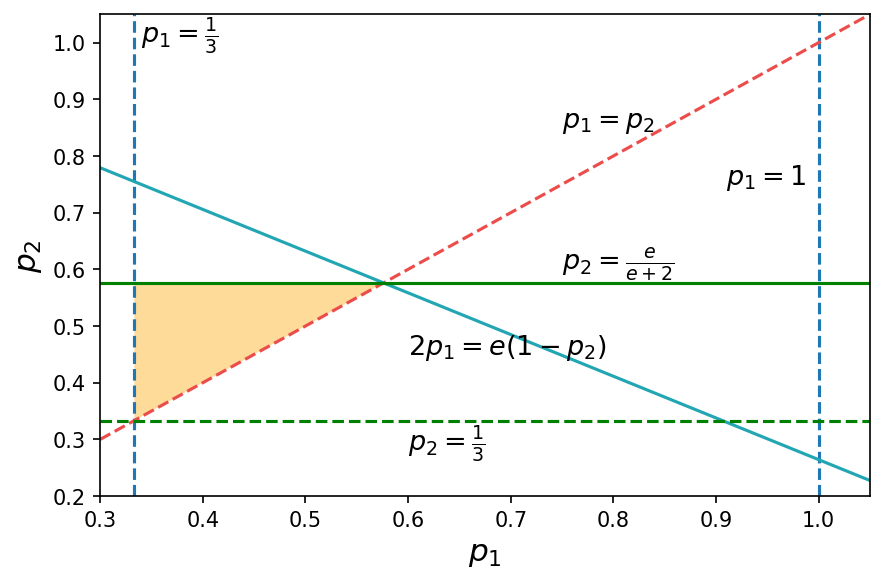}
\end{minipage}
}%
\subfigure[$\Omega^{\dag}_{3}(2)$]{
\begin{minipage}[t]{0.4\linewidth}
\centering
\includegraphics[width=6cm]{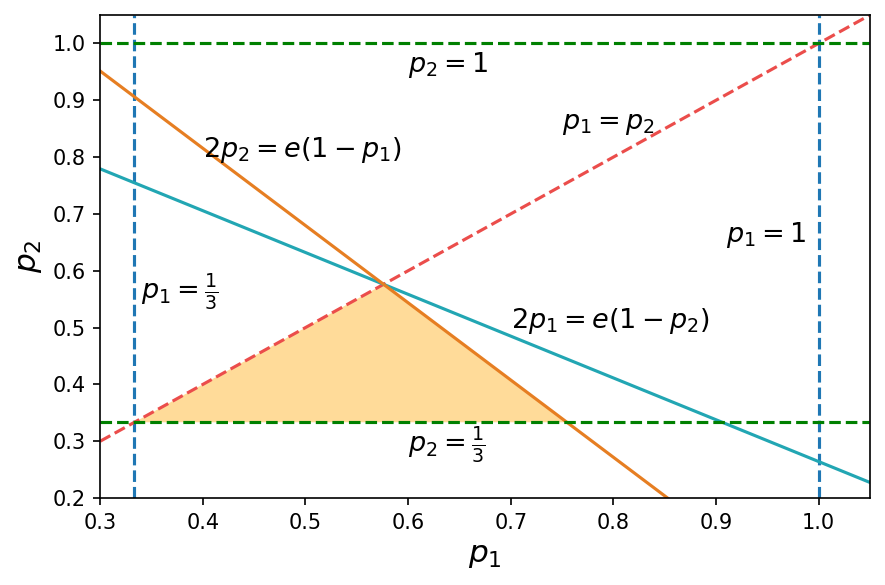}
\end{minipage}
}%
\\
{\small Fig.2. (Color online) The diagram of the feasibility regions $\Omega^{\dag}_{3}(1)$ and $\Omega^{\dag}_{3}(2)$. Panel (a) shows $\Omega^{\dag}_{3}(1)$ and $\Omega^{\dag}_{3}(2)$ is plotted in panel (b).   }
\end{figure*}

Now, our goal is to find

$$\begin{aligned}&\arg\min_{(p_{1},p_{2})\in\Omega^{\dag}_{3}(1)}\text{Var}\left(\widehat{\Pi}(\pi_{0}|\mathbf{P}^{\dag})\right), \quad \text{and}\\
&\arg\min_{(p_{1},p_{2})\in\Omega^{\dag}_{3}(2)}\text{Var}\left(\widehat{\Pi}(\pi_{0}|\mathbf{P}^{\dag})\right)
\end{aligned}$$

\emph{Theorem 4. Given the feasibility regions $\Omega^{\dag}_{3}(1)$ and $\Omega^{\dag}_{3}(2)$ of RR$^{\dag}_{3}$, we have}

\begin{subequations}
\label{eq:whole}
\begin{eqnarray}
\arg\min_{(p_{1},p_{2})\in\Omega^{\dag}_{3}(1)}\text{Var}\left(\widehat{\Pi}(\pi_{0}|\mathbf{P}^{\dag})\right)=\emptyset,\label{subeq:MF-4-1-2-1-1}
\end{eqnarray}
\begin{eqnarray}
\arg\min_{(p_{1},p_{2})\in\Omega^{\dag}_{3}(2)}\text{Var}\left(\widehat{\Pi}(\pi_{0}|\mathbf{P}^{\dag})\right)=\emptyset.\label{subeq:MF-4-1-2-1-2}
\end{eqnarray}
\end{subequations}

\emph{Proof} Because there are two variables $p_{1}$ and $p_{2}$ in $\Omega^{\dag}_{3}$, it suffices to perform partial derivative on $\text{Var}\left(\widehat{\Pi}(\pi_{0}|\mathbf{P}^{\dag})\right)$ with respect to each variable, i.e.,

\begin{subequations}
\label{eq:whole}
\begin{eqnarray}
\begin{aligned}\widehat{\Pi}^{\dag}_{1}&=\frac{\partial\text{Var}\left(\widehat{\Pi}(\pi_{0}|\mathbf{P}^{\dag})\right)}{\partial p_{1}}\\
&=\frac{-4(1-p^{2}_{2})-4p_{2}(2p_{1}+p_{2}-1)\pi_{0}}{(2p_{1}+p_{2}-1)^{3}N}
\end{aligned},\label{subeq:MF-4-1-2-2-1}
\end{eqnarray}
\begin{eqnarray}
\begin{aligned}\widehat{\Pi}^{\dag}_{2}&=\frac{\partial\text{Var}\left(\widehat{\Pi}(\pi_{0}|\mathbf{P}^{\dag})\right)}{\partial p_{2}}\\
&=\frac{-2(1-p^{2}_{2})-[2p_{2}+(1-4p_{1})\pi_{0}](2p_{1}+p_{2}-1)}{(2p_{1}+p_{2}-1)^{3}N}
\end{aligned}.\label{subeq:MF-4-1-2-2-2}
\end{eqnarray}
\end{subequations}
From which we certainly find $\widehat{\Pi}^{\dag}_{1}<0$. In the following, we consider two distinct cases in order to verify the correctness of Theorem 4.

\textbf{Case 1} Considering the feasibility region $\Omega^{\dag}_{3}(1)$, from Eq.(\ref{subeq:MF-4-1-2-2-1}), we see that if there exist the desirable pairs $(p^{\dag}_{1},p^{\dag}_{2})$, then it is sufficient to only consider each pair $(p_{1},p_{2})$ completely in accord with equality $p_{1}=p_{2}$. However, as shown in panel (a) of Fig.2, arbitrary feasible solution $(p_{1},p_{2})$ does not belong to the line segment $p_{1}-p_{2}=0$ at the boundary of the feasibility region $\Omega^{\dag}_{3}(1)$. This completes the proof of Eq.(\ref{subeq:MF-4-1-2-1-1}).

\textbf{Case 2} For the feasibility region $\Omega^{\dag}_{3}(2)$, a similar analysis also holds true. That is to say, If there exist pairs $(p^{\dag}_{1},p^{\dag}_{2})$ we are seeking for, they are only observed on the line segment at the boundary of the feasibility region $\Omega^{\dag}_{3}(2)$, i.e., $(p^{\dag}_{1},p^{\dag}_{2})$ belonging to set $\Lambda$,

$$\Lambda:=\arg\min_{\frac{1}{3}<p_{1}\leq\frac{e^{\epsilon}}{3}<1}\lim_{p_{2}\rightarrow (\frac{1}{3})^{+}}\text{Var}\left(\widehat{\Pi}(\pi_{0}|\mathbf{P}^{\dag})\right).$$

From now on, we begin by determining set $\Lambda$. Note that it is meaningless for function $\text{Var}\left(\widehat{\Pi}(\pi_{0}|\mathbf{P}^{\dag})\right)$ when $2p_{1}+p_{2}-1=0$. In particular, if $p_{2}$ is equal to $\frac{1}{3}$, then $p_{1}$ can not be $\frac{1}{3}$. Accordingly, set $\Lambda$ is rewritten as

$$\Lambda:=\arg\min_{\frac{1}{3}<p_{1}\leq\frac{e^{\epsilon}}{3}<1}\text{Var}\left(\widehat{\Pi}(\pi_{0}|\mathbf{P}^{\dag})\right)_{p_{2}=\frac{1}{3}}.$$

Substituting $p_{2}=\frac{1}{3}$ into Eq.(\ref{eqa:MF-3-3-111}) yields

\begin{equation}\label{eqa:MF-4-1-2-3}
\text{Var}\left(\widehat{\Pi}(\pi_{0}|\mathbf{P}^{\dag})\right)_{p_{2}=\frac{1}{3}}=\frac{\frac{8}{9}+\frac{2}{3}\left(2p_{1}-\frac{2}{3}\right)\pi_{0}}{\left(2p_{1}-\frac{2}{3}\right)^{2}N}-\frac{\pi^{2}_{0}}{N}.
\end{equation}
Taking derivative on both hand sides of Eq.(\ref{eqa:MF-4-1-2-3}) with respect to $p_{1}$, we obtain

\begin{equation}\label{eqa:MF-4-1-2-4}
\begin{aligned}&\frac{\partial\text{Var}\left(\widehat{\Pi}(\pi_{0}|\mathbf{P}^{\dag})\right)_{p_{2}=\frac{1}{3}}}{\partial p_{1}}\\
&=\frac{\frac{4}{3}p_{1}\left(2p_{1}-\frac{2}{3}\right)\pi_{0}-4\left[\frac{8}{9}+\frac{2}{3}\left(2p_{1}-\frac{2}{3}\right)\pi_{0}\right]}{\left(2p_{1}-\frac{2}{3}\right)^{3}N}\\
&<0
\end{aligned},
\end{equation}
which means

$$\Lambda:=\arg\min_{\frac{1}{3}<p_{1}\leq\frac{e^{\epsilon}}{3}<1}\text{Var}\left(\widehat{\Pi}(\pi_{0}|\mathbf{P}^{\dag})\right)_{p_{2}=\frac{1}{3}}=\left\{\frac{e^{\epsilon}}{3}\right\}.$$
To sum up, we complete the proof of Theorem 4. \qed

As mentioned previously, other cases regarding to RR$^{\dag}_{3}$ are omitted. At the same time, the detailed analyses corresponding to RR$_{3}^{\ddag}$ and RR$_{3}^{\clubsuit}$ are omitted as well. Yet, the relevant applications upon RR$_{3}^{\ddag}$ and RR$_{3}^{\clubsuit}$ are investigated in depth in the next section.

\subsection{Applications}

From here on out, we elaborate on some practical applications upon the proposed EWRR$_{3}$, RR$^{\dag}_{3}$, RR$_{3}^{\ddag}$ and RR$_{3}^{\clubsuit}$ to privacy preserving data analysis. In general, these RR-based schemes can be applied into a great variety of scenarios. Particularly, below we take weighted bipartite graph as an example to show more details. In the meantime, we are concerned with applications in the LDP setting. A similar analysis is also suitable for the DP setting, yet we omit it here for brevity.

\subsubsection{EWRR$_{3}$ and RR$^{\dag}_{3}$}

First of all, let us recall the concrete problem stated in Def.5. For a weighted bipartite graph $\mathcal{K}_{\mathcal{V}_{1},\mathcal{V}_{2}}$, suppose that each vertex $v^{i}_{1}$ ($i\in\{1,2,\dots,n\}$) in set $\mathcal{V}_{1}$ represents an individual participant, and a participant $v^{i}_{1}$ is initially connected to $s_{i}$ vertices $v^{j}_{2}$ in the other vertex set $\mathcal{V}_{1}$. At the same time, each vertex $v^{j}_{2}$ ($j\in\{1,2,\dots,m\}$) in set $\mathcal{V}_{2}$ is assumed to be an event. As a result, a number $n$ of participants are grouped int set $\mathcal{V}_{1}$, and set $\mathcal{V}_{2}$ is composed of all the events of interest. In addition, each edge $v^{i}_{1}v^{j}_{2}$ is assigned one weighted value in set \{$0,0.5,1$\}. For example, in a social network with bipartite underlying structure defined above, weight $0$ of edge $v^{i}_{1}v^{j}_{2}$ indicates that participant $v^{i}_{1}$ has a negative perspective on event $v^{j}_{2}$, weight $0.5$ represents a neutral perspective, and weight $1$ corresponds to a positive perspective. For brevity, we also assume that each participant initially selects the most interested event and then joins in survey. In other words, each vertex $v^{i}_{1}$ is just connected to a unique vertex $v^{j}_{2}$. In the jargon of graph theory, the degree $k_{v^{i}_{1}}$ of each vertex $v^{i}_{1}$ is equal to $1$ in the underlying simple graph. The case of $k_{v^{i}_{1}}\geq 1$ is deferred to discuss in next subsections.

In a survey, the data collector wants to evaluate the weight $w_{v^{j}_{2}}$ of each vertex $v^{j}_{2}$  in set $\mathcal{V}_{2}$ in a differentially private manner. Towards this end, each participant $v^{i}_{1}$ makes use of EWRR$_{3}$ or RR$^{\dag}_{3}$ to send a sanitized answer to the data collector. Here, we take EWRR$_{3}$ as an example. Specifically, participant $v^{i}_{1}$ sends the genuine answer to the data collector with probability $p$, or sends either other answer with probability $(1-p)/2$. From Eqs.(\ref{eqa:MF-3-3-110}) and (\ref{eqa:MF-4-1-1}), one see that this kind of approach meets the requirement, i.e., achieving LDP. Meanwhile, the data collector has the ability to obtain unbiased estimation for each quantity $w_{v^{j}_{2}}$ that he cares about. More importantly, the optimal approaches have been shown in Eqs.(\ref{eqa:MF-4-1-1}) and (\ref{eqa:MF-4-1-3}), and thus can be adopted to not only obtain the desirable results but also minimize the errors from statistics itself.

\subsubsection{RR$_{3}^{\ddag}$}

In the preceding subsection, we focus mainly on the privacy preserving analysis on weighted bipartite graph $\mathcal{K}_{\mathcal{V}_{1},\mathcal{V}_{2}}$ in one-dimensional parameter scenario. Namely, the weighted value is a unique parameter. Beyond that, each vertex $v^{i}_{1}$ is assumed to only connect to a vertex $v^{j}_{2}$. Here, we will study a more general problem as follows.

Along the same research line in subsection IV.B.1, we now endow the object, weighted bipartite graph $\mathcal{K}_{\mathcal{V}_{1},\mathcal{V}_{2}}$, with richer properties. Specifically, (1) each vertex $v^{i}_{1}$ is indeed linked with $s_{i}$ vertices $v^{j}_{2}$, and (2) each edge $v^{i}_{1}v^{j}_{2}$ has a weighted value in [$-1,1$]. In this sense, the current goal of date collector is to estimate two parameters of interest, degree $k_{v^{j}_{2}}$ of each vertex $v^{j}_{2}$ and the corresponding weighted value $w_{v^{j}_{2}}$. The associated expressions are given in Eqs.(\ref{eqa:MF-2-4}) and (\ref{eqa:MF-2-5}).

First, different from the case in subsection IV.B.1, the weighted value $w(e_{v^{i}_{1}v^{j}_{2}})$ is an arbitrary real number in range [$-1,1$] in the present setting. Facing with this case, it is necessary to discretize the numerical value before adopting LDP perturbation scheme. The commonly used approach to this kind of task is \emph{Harmony} firstly proposed by Nguyen \emph{et al} \cite{Nguyen-2016}. The heart of \emph{Harmony} is to discretize a numerical value into a binary value and then to perturb it by using two-elements RR. In a follow-up paper \cite{Ye-2019}, Ye \emph{et al} indeed utilized \emph{Harmony} as an important ingredient to propose \emph{PrivKVM} for analyzing Key-Value pair in a LDP way. For convenience, we briefly introduce a delicate version regarding to \emph{Harmony} used in \emph{PrivKVM}, called $VPP$, which is shown in Algorithm 1. With $VPP$, Ye \emph{et al} presented \emph{LPP}. And, for more details to see Ref.\cite{Ye-2019}.

\begin{algorithm}
    	\caption{\emph{VPP}}
	\label{alg:Framwork}
	\begin{algorithmic}[1]
		\Require
		Set $W^{i}_{1}\in[-1,1]^{m}$ composed of edge weights $w(e_{v^{i}_{1}v^{j}_{2}})$ of participant $v^{i}_{1}$;

       \quad Privacy budget $\epsilon_{2}$
		\Ensure
		Perturbed set $W^{i\star}_{1}$
		\State
		Let $W^{i\star}_{1}=\langle0,0,\cdots,0\rangle$
        \State
		Sample $j$ uniformly at random from $\{1,2,\cdots,m\}$, set $w=w(e_{v^{i}_{1}v^{j}_{2}})$
		\State
		Discretization:
        $w^{\star}=\left\{\begin{aligned}&1\qquad \text{w.p.}\quad \frac{w+1}{2}\\
        &-1\quad \text{w.p.}\quad \frac{1-w}{2}
        \end{aligned}\right.$
        \State
        Perturbation: $w^{\star}=\left\{\begin{aligned}&w^{\star}\qquad \text{w.p.}\quad \frac{e^{\epsilon_{2}}}{e^{\epsilon_{2}}+1}\\
        &-w^{\star}\quad \text{w.p.}\quad \frac{1}{e^{\epsilon_{2}}+1}
        \end{aligned}\right.$
		\State
		Return $w^{\star}$
	\end{algorithmic}
\end{algorithm}

With the notation defined above, we reorganize \emph{LPP} in order to make the following analysis more self-contained. Specifically, a participant $v^{i}_{1}$ first selects uniformly an index $j$ in set $\{1,2,\cdots,m\}$, then decides whether there is an edge $e_{v^{i}_{1}v^{j}_{2}}$ or not. If edge $e_{v^{i}_{1}v^{j}_{2}}$ exists, then he adopts $VPP$ to perturb edge weighted $w(e_{v^{i}_{1}v^{j}_{2}})$ to produce $w^{\star}(e_{v^{i}_{1}v^{j}_{2}})$, next sends either a tuple $(1,w^{\star}(e_{v^{i}_{1}v^{j}_{2}}))$ with probability $e^{\epsilon_{1}}/(e^{\epsilon_{1}}+1)$, or a tuple $(0,0)$ with probability $1/(e^{\epsilon_{1}}+1)$. Else, he randomly draws a value $\widetilde{w}$ from in set [$-1,1$], then again executes $VPP$ on weight $\widetilde{w}$ to yield $\widetilde{w}^{\star}$, next sends either a tuple $(1,\widetilde{w}^{\star})$ with probability $1/(e^{\epsilon_{1}}+1)$, or a tuple $(0,0)$ with probability $e^{\epsilon_{1}}/(e^{\epsilon_{1}}+1)$. From which we find that there are three distinct tuples, i.e., $(0,0)$, $(1,1)$ and $(1,-1)$, observed on results output by \emph{LPP}. Based on this, the probability matrix $\mathbf{P}_{3\times 3}(LPP)$ behind \emph{LPP} is expressed in Eq.(\ref{eqa:MF-4-2-2-0}) 
\begin{equation}\label{eqa:MF-4-2-2-0}
\mathbf{P}_{3\times 3}(LPP)=\left(
               \begin{array}{ccc}
                p_{1} &  (1-p_{1})Q_{1}(q,\widetilde{w}) & (1-p_{1})(1-Q_{1}(q,\widetilde{w}))  \\
                   1-p_{1}  &  p_{1}Q_{2}(q,w) & p_{1}(1-Q_{2}(q,w)) \\
                   1-p_{1}  &   p_{1}(1-Q_{2}(q,w))  & p_{1}Q_{2}(q,w) \\
               \end{array}
             \right).
\end{equation}
in which the first row corresponds to tuple $(0,0)$, subsequently followed by tuples $(1,1)$ and $(1,-1)$. We have used symbols

$$\begin{aligned}&p_{1}=\frac{e^{\epsilon_{1}}}{e^{\epsilon_{1}}+1},\quad q=\frac{e^{\epsilon_{2}}}{e^{\epsilon_{2}}+1}, \\
&Q_{1}(q,\widetilde{w})=\frac{(1+\widetilde{w})q}{2}+\frac{(1-\widetilde{w})(1-q)}{2},\\
&Q_{2}(q,w)=\frac{(1+w)q}{2}+\frac{(1-w)(1-q)}{2}.
\end{aligned}$$

As shown in \cite{Ye-2019}, \emph{LPP} satisfies $\epsilon$-LDP where $\epsilon=\epsilon_{1}+\epsilon_{2}$. Here, we also provide a simple proof only in order to make this work more readable.

\emph{Lemma 1 \emph{LPP} satisfies $\epsilon$-LDP where $\epsilon=\epsilon_{1}+\epsilon_{2}$.}

\emph{Proof} By Def.1, it is straightforward to obtain Eq.(\ref{eqa:MF-4-2-2-1}).

\begin{equation}\label{eqa:MF-4-2-2-1}
e^{\epsilon}=\max\left\{\max_{\widetilde{w}, w\in [-1,1]}\left\{\frac{p_{1}}{1-p_{1}}, \frac{p_{1}Q_{2}(q,w)}{(1-p_{1})Q_{1}(q,\widetilde{w})},\frac{p_{1}(1-Q_{2}(q,w))}{(1-p_{1})Q_{1}(q,\widetilde{w})}\right\}\right\}.
\end{equation}
This is equivalently translated into determining the following expression
\begin{equation}\label{eqa:MF-4-2-2-2}
\begin{aligned}e^{\epsilon-\epsilon_{1}}&=\max\left\{\max_{\widetilde{w}, w\in [-1,1]}\left\{ \frac{Q_{2}(q,w)}{Q_{1}(q,\widetilde{w})},\frac{1-Q_{2}(q,w)}{Q_{1}(q,\widetilde{w})}\right\}\right\}\\
&=\max\left\{\frac{\max_{ w\in [-1,1]}\{Q_{2}(q,w), 1-Q_{2}(q,w)\}}{\min_{ \widetilde{w}\in [-1,1]}\{Q_{1}(q,\widetilde{w})\}}\right\}
\end{aligned}.
\end{equation}
It is easy to see that $Q_{2}(q,w)$ is an increasing function with respect to variable $w$, and $Q_{1}(q,\widetilde{w})$ is also an increasing function of variable $\widetilde{w}$. Thus, we have
\begin{equation}\label{eqa:MF-4-2-2-3}
e^{\epsilon-\epsilon_{1}}=\frac{Q_{2}(q,1)}{Q_{1}(q,-1)}=e^{\epsilon_{2}},
\end{equation}
and obtain $\epsilon=\epsilon_{1}+\epsilon_{2}$. This means that \emph{LPP} indeed satisfies $\epsilon$-LDP as desired. \qed

As we can see above, there are two-elements RRs used to develop \emph{LPP}. For a given privacy budget $\epsilon$($=\epsilon_{1}+\epsilon_{2}$), if $\epsilon_{2}$ is assigned for two-elements RRs associated with $VPP$ and $\epsilon_{1}$ is deployed into the other as manipulated in scheme \emph{LPP}, then the resultant scheme (in fact, \emph{LPP}) achieves the theoretical upper bound for the guarantee of differential privacy as mentioned in Remark 2. It is natural to ask that under the situation above, whether or not there exists other protocol that supports much lower bound for privacy budget. To answer this problem, let we divert insight into Eq.(\ref{eqa:MF-4-2-2-2}), and observe that what we need do is to minimize  term on the right hand side of the last equality. To make further progress, this task is addressed by minimizing term $Q_{2}(q,w)$ or maximizing term $Q_{1}(q,\widetilde{w})$, or both. Note that we require that $Q_{2}(q,w)$ be more than $1/2$. If no, we should consider term $1-Q_{2}(q,w)$. By symmetry, $Q_{2}(q,w)$ is assumed to be larger than $1/2$. Next, as discussed previously, weight $w$ may take an arbitrary value in range [$-1,1$] and $Q_{2}(q,w)$ is an increasing function of $w$, hence, we have no ability to minimize function $Q_{2}(q,w)$ under the constraint of LDP and are forced to admit $\max_{ w\in [-1,1]}\{Q_{2}(q,w)\}=Q_{2}(q,1)$. As an immediate consequence, we only appeal to maximizing term $Q_{1}(q,\widetilde{w})$. In view of assumption above, weight $\widetilde{w}$ is a variable drawn uniformly at random from range [$-1,1$], the corresponding expectation $E[\widetilde{w}]$ is equal to zero by virtue of the Large Number Theorem. Therefore, this kind of choice of $\widetilde{w}$ has no significant influence on estimations of weight $w_{v^{j}_{2}}$. In theory, an arbitrary choice of $\widetilde{w}$ is admitted. However, some of them introduce more noisy into estimation of quantities we are interested in, and, simultaneously, lead to more complicated computations before we obtain the unbiased estimators for quantities. According to the statement above, for convenience, we select a simplest choice manner where the pending weight $\widetilde{w}$ is always assumed to be $0$ when edge $e_{v^{i}_{1}v^{j}_{2}}$ does not exist. In this case, the probability matrix is of form

$$\mathbf{P}^{\ddag}_{3\times 3}(LPP)=\left(
               \begin{array}{ccc}
                p_{1} &  (1-p_{1})/2 & (1-p_{1})/2  \\
                   1-p_{1}  &  p_{1}p_{2} & p_{1}(1-p_{2})  \\
                   1-p_{1} &  p_{1}(1-p_{2})  & p_{1}p_{2} \\
               \end{array}
             \right),$$
in which
$$p_{1}=\frac{e^{\epsilon_{1}}}{e^{\epsilon_{1}}+1},\quad q=\frac{e^{\epsilon_{2}}}{e^{\epsilon_{2}}+1}, \quad p_{2}=\frac{(1+w)q}{2}+\frac{(1-w)(1-q)}{2}.$$
This is a version of three-elements RR that we have studied in previous sections. With Corollary 4, the scheme, denoted by $LPP^{\ddag}$, outputs unbiased estimations for degree $k_{v^{j}_{2}}$ and weight $w_{v^{j}_{2}}$ of vertex $v^{j}_{2}$ in set $\mathcal{V}_{2}$.
In the meantime, the LDP property of $LPP^{\ddag}$ is easily verified, and shown in the following lemma.

\emph{Lemma 2 $LPP^{\ddag}$ satisfies $\epsilon'$-LDP, and the precise solution of $\epsilon$ is given by}

\begin{equation}\label{eqa:MF-4-2-2-4}
\epsilon'=\ln\left(\frac{2e^{\epsilon_{1}+\epsilon_{2}}}{e^{\epsilon_{2}}+1}\right).
\end{equation}

\emph{Proof} Along the similar research line as in proving Lemma 1, we need to determine

\begin{equation}\label{eqa:MF-4-2-2-5}
e^{\epsilon'}=\max\left\{\max_{w\in [-1,1]}\left\{\frac{p_{1}}{1-p_{1}}, \frac{p_{1}p_{2}}{(1-p_{1})/2},\frac{p_{1}(1-p_{2})}{(1-p_{1})/2}\right\}\right\}.
\end{equation}
After some algebra, it is straightforward to obtain

\begin{equation}\label{eqa:MF-4-2-2-6}
e^{\epsilon'-\epsilon_{1}}=\max\left\{\max_{w\in [-1,1]}\left\{1, 2p_{2}, 2(1-p_{2})\right\}\right\}=\frac{2e^{\epsilon_{2}}}{e^{\epsilon_{2}}+1}.
\end{equation}
From which we derive

\begin{equation}\label{eqa:MF-4-2-2-6}
e^{\epsilon'}=\frac{2e^{\epsilon_{1}+\epsilon_{2}}}{e^{\epsilon_{2}}+1}.
\end{equation}
This completes the proof of Lemma 2. \qed

With Eqs.(\ref{eqa:MF-4-2-2-3}) and (\ref{eqa:MF-4-2-2-6}), we conclude that the proposed scheme $LPP^{\ddag}$ is more optimal compared to the previous $LPP$ because the privacy budget ``output"\footnote[3]{For a given LDP protocol $\mathcal{M}$, one takes a predefined privacy budget $\epsilon$ as input of $\mathcal{M}$. After that, $\mathcal{M}$ turns out to be $\epsilon'$-LDP. For brevity, the latter $\epsilon'$ is seemed as privacy budget output by $\mathcal{M}$. Strictly speaking, $\epsilon'$ is a product after running $\mathcal{M}$ from the algorithm point of view. In general, $\epsilon$ is equal to $\epsilon'$. However, it is also likely to encounter case of $\epsilon\neq\epsilon'$ in many LDP $\mathcal{M}$s, for instance, scheme $LPP^{\ddag}$.} by the former is lower than that in the latter under the same scenario, namely, when consuming the same amount of privacy budget. In other words, the former provides much stronger guarantee for privacy than the latter given a privacy budget $\epsilon$. Besides that, the proposed $LPP^{\ddag}$ achieves optimal bound for the output privacy budget. Specifically, more details are shown in Lemma 3.

\emph{Lemma 3 Given the probability matrix $\mathbf{P}^{\ddag}_{3\times 3}(LPP)$, $LPP^{\ddag}$ achieves optimal bound for the output privacy budget, namely,}

\begin{equation}\label{eqa:MF-4-2-2-7}
\epsilon'=\epsilon^{\dagger}=\text{inf}\{\epsilon_{j}^{\star}|j\in\mathfrak{J}\},
\end{equation}
\emph{where $\mathfrak{J}$ is an index set that consists of all the indices corresponding to LDP protocols built upon probability matrix $\mathbf{P}_{3\times 3}(LPP)$.}

Before beginning our demonstrations, it should be pointed out that in the next proof, we use many results derived in Lemmas 1 and 2 directly for the sake of simplicity and readability. The corresponding details are clear to understand with the help of Lemmas 1 and 2.

\emph{Proof} From definition of probability matrix $\mathbf{P}^{\ddag}_{3\times 3}(LPP)$ and the symmetric property shown in the second and third columns, it remains to determine

\begin{equation}\label{eqa:MF-4-2-2-8}
\arg\min\max_{\widetilde{w}\in [-1,1]}\left\{ \frac{Q_{2}(q,1)}{Q_{1}(q,\widetilde{w})}, \frac{Q_{2}(q,1)}{1-Q_{1}(q,\widetilde{w})}\right\}.
\end{equation}
According to the concrete expression of $Q_{1}(q,\widetilde{w})$, we obtain

\begin{equation}\label{eqa:MF-4-2-2-9}
\begin{aligned}f(\widetilde{w}):&=\max_{\widetilde{w}\in [-1,1]}\left\{ \frac{Q_{2}(q,1)}{Q_{1}(q,\widetilde{w})}, \frac{Q_{2}(q,1)}{1-Q_{1}(q,\widetilde{w})}\right\}\\
&=\left\{\begin{aligned}&\frac{Q_{2}(q,1)}{Q_{1}(q,\widetilde{w})}, \quad \widetilde{w}\in[-1,0]\\
&\frac{Q_{2}(q,1)}{1-Q_{1}(q,\widetilde{w})},\quad \widetilde{w}\in[0,1]
\end{aligned}\right.
\end{aligned}.
\end{equation}
Therefore, we have
\begin{equation}\label{eqa:MF-4-2-2-10}
\arg\min f(\widetilde{w})=\{0\},
\end{equation}
which implies
\begin{equation}\label{eqa:MF-4-2-2-11}
e^{\epsilon'-\epsilon_{1}}=\frac{Q_{2}(q,1)}{Q_{1}(q,0)}=\frac{2e^{\epsilon_{2}}}{e^{\epsilon_{2}}+1}.
\end{equation}
This means that $LPP^{\ddag}$ indeed achieves optimal bound for the output privacy budget, i.e., $e^{\epsilon'}=\frac{2e^{\epsilon_{1}+\epsilon_{2}}}{e^{\epsilon_{2}}+1}$. We prove the correctness of Lemma 3. \qed

At last, it is worth noticing that while the ratio of two unbiased estimators obtained from $LPP^{\ddag}$ is not usually unbiased, the corresponding consistence is guaranteed in terms of the Large Number Theorem. That is to say, the ratio is gradually close to the truth value when increasing the number of participants.

\subsubsection{RR$_{3}^{\clubsuit}$}

The goal of this subsection is to show many applications of three-elements RR$_{3}^{\clubsuit}$. As previously, we still select a weighted bipartite graph $\mathcal{K}_{\mathcal{V}_{1},\mathcal{V}_{2}}$ as an example application. In the meantime, what we are going to discuss is the same problem as defined in subsection IV.B.2.

First of all, let us briefly introduce a published protocol, called \emph{PCKV-UE}, due to Gu \emph{et al} \cite{Gu-2020}. In fact, protocol \emph{PCKV-UE} is one follow-up work for $LPP$. Nonetheless, this protocol is established based on two different ingredients, Padding-and-Sampling and Unary Encoding, from that used in $LPP$. Below, we mainly provide the associated probability matrix, denoted by $\mathbf{P}^{\clubsuit}_{3\times 3}(UE)$, and more details about how to manipulate \emph{PCKV-UE} to obtain estimator for quantities of interest are omitted. For more details to see Ref.\cite{Gu-2020}.

Below is a non-rigorous yet understandable description about \emph{PCKV-UE}. The first step in \emph{PCKV-UE} is to discretize the weight $w(e_{v^{i}_{1}v^{j}_{2}})$ of each existing edge $e_{v^{i}_{1}v^{j}_{2}}$ of graph $\mathcal{K}_{\mathcal{V}_{1},\mathcal{V}_{2}}$ in a similar way as used in \emph{VPP}. Note that we continue to denote by  $w^{\star}(e_{v^{i}_{1}v^{j}_{2}})$ the perturbed edge weight. Then, let $w^{\star}(e_{v^{i}_{1}v^{j}_{2}})$ be equal to either $1$ with probability $(1+w(e_{v^{i}_{1}v^{j}_{2}}))/2$, or $-1$ with probability $(1-w(e_{v^{i}_{1}v^{j}_{2}}))/2$. In addition, we do nothing for an arbitrary non-existing edge. After that, there is always a unique tuple in set $\{(1,1), (1,-1), (0,0)\}$ linked with each vertex pair where one vertex is from set $\mathcal{V}_{1}$ and the other from set $\mathcal{V}_{2}$. For example, tuple $(0,0)$ indicates that there is no edge between vertex pair in question. Next, \emph{PCKV-UE} perturbs each tuple in a slightly different fashion from that used in \emph{LPP}. Specifically, given a tuple ($\alpha,\beta$), (1) if $\alpha$ and $\beta$ are equal to $1$, then this tuple either remains unchanged with probability $ap$, or is set to $(1,-1)$ with probability $a(1-p)$, or is switched into tuple $(0,0)$ with probability $1-a$; (2) if $\alpha$ is equal to $1$ and $\beta$ is equal to $-1$, then this tuple either remains unchanged with probability $ap$, or is set to $(1,1)$ with probability $a(1-p)$, or is switched into tuple $(0,0)$ with probability $1-a$; (3) if $\alpha$ and $\beta$ are equivalent to $0$, then this tuple either remains unchanged with probability $b$, or is set to either $(1,1)$ or $(1,-1)$ with an identical probability $b/2$. From which we express the probability matrix $\mathbf{P}^{\clubsuit}_{3\times 3}(UE)$ in the next form

$$\mathbf{P}^{\clubsuit}_{3\times 3}(UE)=\left(
               \begin{array}{ccc}
                b &  (1-b)/2 & (1-b)/2  \\
                   1-a  &  ap & a(1-p)  \\
                   1-a &  a(1-p)  & ap \\
               \end{array}
             \right).$$
Note that there are predefined relationships between entries in matrix $\mathbf{P}^{\clubsuit}_{3\times 3}(UE)$ (see \cite{Gu-2020} for details), which are shown as follows
$$\frac{ab}{(1-a)(1-b)}=e^{\epsilon_{1}}, \quad  p=\frac{e^{\epsilon_{2}}}{e^{\epsilon_{2}}+1}, \quad a,b,p\in[1/2,1).$$
It is noteworthy that the result output by protocol $\emph{PCKV-UE}$ is no longer a tuple but a string $\textbf{y}$ in which only an position is equal to $1$ and all other positions are assigned with $0$. For instance, for a given tuple $(1,1)$ with index $j$ \footnote[4]{The index $j$ corresponding to a tuple $(\alpha,\beta)$ hints that edge we are discussing has one endpoint $v^{j}_{2}$. At the same time, we use $\textbf{y}_{j}$ to denote $j$th standard basis vector.}, the probability of finding $\textbf{y}_{j}=[0,\cdots,1,\cdots,0]$ is equal to $ap$.

Obviously, the matrix $\mathbf{P}^{\clubsuit}_{3\times 3}(UE)$ is a simple version of matrix $\mathbf{P}^{\clubsuit}_{3\times 3}$ introduced in subsection III.C. In view of Corollary 5, we conclude that the estimations for degree $k_{v^{j}_{2}}$ and weight $w_{v^{j}_{2}}$ of each vertex $v^{j}_{2}$ in set $\mathcal{V}_{2}$ are unbiased. While the $\epsilon$-LDP property of protocol \emph{PCKV-UE} has been verified in \cite{Gu-2020}, we provide a brief proof for the purpose of making this work more self-contained.

\emph{Lemma 4 $\mathbf{P}^{\clubsuit}_{3\times 3}(UE)$ follows $\epsilon$-LDP and the precise solution of $\epsilon$ is given by }

\begin{equation}\label{eqa:MF-4-2-3-1}
\epsilon'=\max\left\{\epsilon_{2},\ln\left(\frac{2e^{\epsilon_{1}+\epsilon_{2}}}{e^{\epsilon_{2}}+1}\right)\right\}.
\end{equation}

\emph{Proof} By Def.1 and probability matrix $\mathbf{P}^{\clubsuit}_{3\times 3}(UE)$, it suffices to measure

\begin{equation}\label{eqa:MF-4-2-3-2}
\begin{aligned}\epsilon&=\max\left\{ \ln\left(\frac{ap}{(1-b)/2}\times\frac{a}{1-b}\right), \frac{ap}{a(1-p)}\right\}\\
&=\max\left\{\epsilon_{2},\ln\left(\frac{2e^{\epsilon_{1}+\epsilon_{2}}}{e^{\epsilon_{2}}+1}\right)\right\}
\end{aligned}.
\end{equation}
This completes the proof of Lemma 4. \qed

It should be mentioned that we have used an intrinsic property of Unary Encoding to consolidate the proof above. Specifically, for a given pair of distinct tuples $(\alpha_{1},\beta_{1})$ with index $i$ and $(\alpha_{2},\beta_{2})$ with index $j$, (1) if $i=j$ holds, we need to only perturb $\beta_{s}$ ($s=1,2$) to yield an identical value $\beta$ in $\{0,-1,1\}$, then the probability we are trying to seek is $\frac{ap}{a(1-p)}$; (2) if $i\neq j$ holds, then the probability is $\frac{ap}{(1-b)/2}\times\frac{a}{1-b}$  because we now need to perturb entries on two positions $i$ and $j$ in each string in order to produce an identical string. Note also that the problem of dividing a given privacy budget $\epsilon$ using result in Eq.(\ref{eqa:MF-4-2-3-1}) into two available portions, each for one perturbation step in protocol $\emph{PCKV-UE}$, is out of scope of this work. Thus, we omit it here. A near-optimal allocation mechanism for privacy budget $\epsilon$ has been studied in \cite{Gu-2020}.

\section{Related work}

Privacy data analysis has given rise to great interest over the recent years due to various kinds of applications \cite{Salloum-2019,Cuzzocrea-2018}. Meanwhile, it is well known that the collected data often contain many private information, such as gender, IDs, phone number, etc. Analyzing data without any security guarantee leads to private information leakage \cite{Guo-2020}-\cite{Goldfeld-2017}, which further causes that data owners are reluctant to share their own data. Therefore, the need increases for a robust, meaningful, and mathematically rigorous definition of privacy, together with a computationally rich class of algorithms that satisfy this definition. Differential privacy is such a definition \cite{Dwork-2014}.

Since Dwork first proposed differential privacy in 2006 \cite{Dwork-2006}, the related research has received more attention \cite{Dwork-2014}. As a result, there are many approaches developed to achieve differential privacy, for instance, Laplacian mechanism \cite{Dwork-2006-1}, exponential mechanism \cite{Dwork-2014}, randomized response \cite{Erlingsson-2014}, etc. Among which, randomized response has been widely adopted into a great number of algorithms satisfying differential privacy, such as RAPPOR \cite{Erlingsson-2014}, k-RR \cite{Kairouz-2014}, O-RR \cite{Kairouz-2016}, Harmony-mean \cite{Nguyen-2016}, SHist \cite{Bassily-2015}, PrivKVM \cite{Ye-2019}, KVUE \cite{Sun-2019} and PCKV \cite{Gu-2020}, by virtue of its own merits including simplicity and understandability. As known, randomized response as a means for collecting statistical information in social science was first presented by Warner in 1965 \cite{Warner-1965}, and is proved to perform well in the estimate of statistics. Since then, this technique has become the state-of-the-art approach for protecting personal privacy during collecting private data and thus has been popularly used in statistic sciences \cite{Dwork-2014,Ye-2019,Nguyen-2016,Bassily-2015,Sun-2019}.

The original form of randomized response is constructed for applications into a survey with binary answers \cite{Warner-1965}. After that, many researchers have paid substantial attention to investigation on basic properties of the typical randomized response and, in the meantime, also put forward a lot of variants suitable for applications in more general situation, for example, generalized randomized response \cite{Singh-2017}. Studied models include unrelated question model \cite{Greenberg-1968}, Moor's procedure \cite{Moors-1971}, two-stage model \cite{Mangat-1994}, and so on. These models have been widely adopted into many practical scenarios, for instance, corruption \cite{Gingerich-2010}, sexual behavior \cite{Donovan-2003}, faking on a CV \cite{Chen-2014}.

In the context of differential privacy, the relevant properties of randomized response have been studied in detail \cite{Wang-2016,Kairouz-2017,Holohan-2017}. In \cite{Wang-2016}, Wang \emph{et al} examined using randomised response to differentially privately collect data, and also make a comparison of its efficiency with the Laplace mechanism. In \cite{Kairouz-2017}, Kairouz \emph{et al} have shown how to
make use of RR to design optimal differentially private mechanisms for a large group of private multi-party computation problems. In a follow-up work \cite{Holohan-2017}, Holohan \emph{et al} have systematically discussed a generalized randomized response in the area of differential privacy while the corresponding version is of binary answers. Note that the work in \cite{Holohan-2017} focuses on two kinds of versions for randomized response, i.e., strict differential privacy and relaxed differential privacy \cite{Dwork-2014}. Using randomized response with binary answers as an ingredient, Ye \emph{et al} studied Key-Value pair in local differential private setting \cite{Ye-2019}. It should be mentioned that the Key-Value pair is also considered in two recently published papers \cite{Sun-2019,Gu-2020}. As shown previously, this kind of task is in essence closely associated with three-elements randomized response. Outside the focus of this work, Song \emph{et al} have used randomized response to build up multiple sensitive values-oriented personalized privacy preservation \cite{Song-2020}, and Wei \emph{et al} have utilized randomized response to establish local differential private algorithm for collecting and generating decentralized attributed graphs \cite{Wei-2020}.

\section{Conclusion}

In summary, we are mainly concerned with three-elements Randomized Responses, a widely-used technique for differentially private mechanisms, and analyze the relevant quantities of great interest in depth. Taking counting query as an example, we build up a framework for deriving closed-form solutions to expectation and variance of some basic statistic quantities by virtue of Maximal Likelihood Estimate, and, accordingly, obtain the exact solutions of relevant parameters on several fundamental and significant RR mechanisms including EWRR$_{3}$, RR$_{3}^{\dag}$, RR$_{3}^{\ddag}$ and RR$_{3}^{\clubsuit}$. Based on this, we determine the bounds of estimators by minimizing variance, and find the optimal design scheme for EWRR$_{3}$. Furthermore, we show potential applications of EWRR$_{3}$, RR$_{3}^{\dag}$, RR$_{3}^{\ddag}$ and RR$_{3}^{\clubsuit}$ to analyzing weighted bipartite graph in the privacy preserving scenario. We not only discuss some previously published protocols, but also propose optimal scheme achieving tight bound. Last but most importantly, we stress that in the relational data analysis, a portion of privacy budget is sometimes ``consumed" by the proposed LDP mechanism itself accidentally, resulting to a more stronger privacy guarantee than we would get by simply sequential compositions. In this study, we verify this viewpoint by using the analysis of weighted bipartite graph. We firmly believe that there are still many unknown and interesting problems in this field to be solved in the future, which is left as our next move.

\section*{Acknowledgment}

The research was supported by the National Key Research and Development Plan under grant 2020YFB1805400 and the National Natural Science Foundation of China under grant No. 62072010.

\end{document}